\documentclass[11pt]{amsart}
\usepackage[applemac]{inputenc}
\usepackage[titletoc]{appendix}
\usepackage{amssymb}
\usepackage{amsfonts}
\usepackage{amsthm}
\usepackage{color}
\usepackage{graphicx}
\usepackage{xcolor}
\usepackage{shadethm}
\usepackage{subfigure}
\usepackage{epigraph}

\usepackage{enumerate}
\usepackage{verbatim}
\usepackage{lipsum}
\usepackage{tikz}
\usetikzlibrary{trees}
\usetikzlibrary{decorations.markings}
\usetikzlibrary{arrows}
\usepackage{mathrsfs}
\usepackage[margin=1.0in]{geometry}
\usepackage{xparse}

\usetikzlibrary{positioning,arrows,patterns}
\usetikzlibrary{decorations.markings}
\usetikzlibrary{calc}
\tikzset{
  photon/.style={decorate, decoration={snake}, draw=black},
  fermion/.style={draw=black, postaction={decorate},decoration={markings,mark=at position .55 with {\arrow{>}}}},
  vertex/.style={draw,shape=circle,fill=black,minimum size=3pt,inner sep=0pt},
}
\NewDocumentCommand\semiloop{O{black}mmmO{}O{above}}
{%
\draw[#1] let \p1 = ($(#3)-(#2)$) in (#3) arc (#4:({#4+180}):({0.5*veclen(\x1,\y1)})node[midway, #6] {#5};)
}

\usepackage{url}
\usepackage{hyperref}
\hypersetup{colorlinks=false, linktocpage, linkbordercolor=red, citebordercolor=green, filebordercolor=red, urlbordercolor=cyan, pdfborderstyle={/S/U/W 1}}

\theoremstyle{plain}
\newtheorem{thm}{Theorem}[section]

\newtheorem{cor}[thm]{Corollary}

\newtheorem{defn}{Definition}[section]

\theoremstyle{remark}
\newtheorem{rem}{Remark}[section]

\newcommand{\R}{\mathbb{R}}
\newcommand{\A}{\mathcal{A}}

\newcommand{\E}{\mathbb{E}}

\newcommand{\dd}{{\mathrm{d}}}

\newcommand{\id}{\mathrm{id}}

\newcommand{\hateta}{\widehat{\boldsymbol\eta}}
\newcommand{\hatX}{\widehat{\mathsf{X}}}

\newcommand{\de}{\partial}

\newcommand{\calH}{\mathcal{H}}

\newcommand{\calL}{\mathcal{L}}

\newcommand{\calP}{\mathcal{P}}

\def\gpd{\,\lower1pt\hbox{$\longrightarrow$}\hskip-.24in\raise2pt
               \hbox{$\longrightarrow$}\,}

\newcommand{\I}{\mathrm{i}}
\newcommand{\F}{\mathcal{F}}

\newcommand{\ee}{\textnormal{e}}

\begin{document}

\title[RSG Quantization for Constant Poisson Structures]{Relational Symplectic Groupoid Quantization for Constant Poisson Structures}
\author[A. S. Cattaneo]{Alberto S. Cattaneo}
\author[N. Moshayedi]{Nima Moshayedi}
\author[K. Wernli]{Konstantin Wernli}
\address{Institut für Mathematik\\ Universität Zürich\\ 
Winterthurerstrasse 190
CH-8057 Zürich}
\email[A. S.~Cattaneo]{alberto.cattaneo@math.uzh.ch}
\address{Institut für Mathematik\\ Universität Zürich\\ 
Winterthurerstrasse 190
CH-8057 Zürich}
\email[N.~Moshayedi]{nima.moshayedi@math.uzh.ch}
\address{Institut für Mathematik\\ Universität Zürich\\ 
Winterthurerstrasse 190
CH-8057 Zürich}
\email[K.~Wernli]{konstantin.wernli@math.uzh.ch}
\thanks{
This research was (partly) supported by the NCCR SwissMAP, funded by the Swiss National Science Foundation, and by the
COST Action MP1405 QSPACE, supported by COST (European Cooperation in Science and Technology).
}



\maketitle

\begin{abstract}
As a detailed application of the BV-BFV formalism for the quantization of field theories on manifolds with boundary, this note describes a quantization of the relational symplectic groupoid for a constant Poisson structure. The presence of mixed boundary conditions and the globalization of results is also addressed. In particular, the paper includes an extension to space--times with boundary of some formal geometry considerations in the BV-BFV formalism, and specifically introduces into the BV-BFV framework a “differential” version of the classical and quantum master equations.
The quantization constructed in this paper induces Kontsevich’s deformation quantization on the underlying Poisson manifold, i.e., the Moyal product, which is known in full details. This allows focussing on the BV-BFV technology and testing it. For the unexperienced reader, this is also a practical and reasonably simple way to learn it. 
\end{abstract}

\tableofcontents


\section{Introduction}
\begin{figure}[htbp]
   \centering
   \includegraphics[scale=0.8]{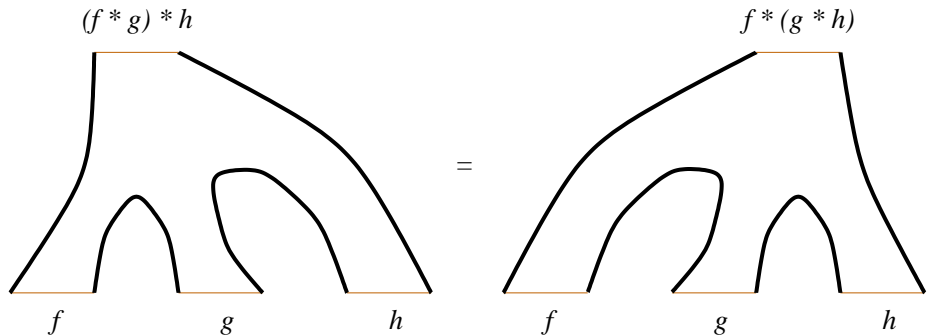} 
   \caption{Associativity in the Poisson sigma model}
   \label{fig:stringassoc}
\end{figure}

Deformation quantization of Poisson manifolds as constructed by Kontsevich (\cite{K}), which
we recall in Section~\ref{s:DQ}, 
corresponds to the
perturbative quantization of the Poisson sigma model (PSM) on a disk with appropriate boundary conditions and boundary observables (\cite{CF1}), see Section~\ref{s:PSM}. 
The associativity of the star product is related to the fact that the PSM is a topological field theory, 
see figure~\ref{fig:stringassoc} for a rough impression. Even though the associativity may be explicitly proved once one has the explicit formulae, as Kontsevich did, it is useful to put its PSM origin on firmer ground.
This is the goal of this note, even though we only focus on the case of constant Poisson structure.

The main idea, sketched in (\cite{CMR2}), is to cut the picture in figure~\ref{fig:stringassoc} and regard its gluing back as the composition of states for the PSM as a quantum field theory. These states turn out to be the quantization of a classical construction (\cite{C,CC1,CC2}) leading to the 
relational symplectic groupoid (RSG), roughly speaking a groupoid in the extended symplectic category, associated to the Poisson manifold. We recap the main facts about the RSG in the first part of Section~\ref{s:RSG}.

For the perturbative construction of the states of the PSM we use the general procedure
introduced by the first author together with P. Mn\"ev and N. Reshetikhin in (\cite{CMR2}) and called
the quantum BV-BFV formalism, which we recall in Section~\ref{s:BVBFV}, where we also rely on the fact that the PSM
is an example of an AKSZ theory (\cite{AKSZ}). For the classical version of the BV-BFV formalism we refer
to \cite{CMR1}, for more details on the BV formalism to (\cite{BV,A,F}), and for another short introduction
of the quantum BV-BFV formalism to (\cite{CMW}).

A crucial point of this quantization is that it first requires a choice of background. In the case of the perturbative PSM, this is just the choice of a constant map or, equivalently, of a point of the target Poisson manifold. To keep track of this choice we use formal geometry as in \cite{BCM}. We recall this construction in Section~\ref{s:BVBFV}. 

In the second
part of Section~\ref{s:RSG} we then apply the formalism to the construction of the states for the PSM with constant Poisson structure.
In Section~\ref{s:mdQME} and ~\ref{s:assoc}
we prove, in the BV-BFV language, that the states are gauge invariant and background independent and
that they define an associative product on the space of physical states. Finally, in Section~\ref{s:main}
we show how the Moyal product is obtained by the composition of states, which in particular justifies
its associativity. In the last section we present an outlook on the case of general Poisson manifolds.



\subsection*{Acknowledgement}
We thank I. Contreras and P. Mn\"ev for useful discussions and comments. We are especially grateful to the referee for pointing out some small errors and for very precious comments and suggestions.

\section{Deformation Quantization}\label{s:DQ}
\subsection{Rough description}

\textsf{Deformation quantization} \cite{BFFLS}
is a quantization procedure that focuses on the observables: the classical algebra of observables (smooth functions on a Poisson manifold) gets \textsf{deformed} into a noncommutative associative algebra.

\begin{defn}[Formal deformation/star product]
A \textsf{formal deformation} (or \textsf{star product}) on an associative algebra $\mathscr{A}$ over a ring $k$ is a $k[[\varepsilon]]$-bilinear map
\[
\star:\mathscr{A}[[\varepsilon]]\times\mathscr{A}[[\varepsilon]]\to \mathscr{A}[[\varepsilon]],
\]
satisfying $\psi\star(\phi\star\xi)=(\psi\star\phi)\star\xi$ for all $\psi,\phi,\xi\in\mathscr{A}[[\varepsilon]]$, i.e. associativity has to hold for $\star$.  If the algebra is unital, one also requires the unit $1$ not to be deformed; i.e., $\psi\star 1=1\star\psi=\psi$ for all $\psi\in \mathscr{A}[[\varepsilon]]$. Moreover, the star product should be a deformation of the original, i.e. for $f,g\in \mathscr{A}$ we require
\[
f\star g=fg+ \sum_{k \geq 1} \varepsilon^kB_k(f,g)\in\mathscr{A}[[\varepsilon]].
\] The product is extended to $\mathscr{A}[[\varepsilon]]$ by $\varepsilon$-bilinearity. 
\end{defn}

\begin{rem} One can show that \[\{f,g\} := \frac{f\star g - g\star f}{\varepsilon}\Big|_{\varepsilon = 0}\] defines a Poisson bracket on $\mathscr{A}$. If $\mathscr{A} = C^{\infty}(M)$ is the algebra of functions on a smooth manifold, we require the operators $B_k$ to be bidifferential operators.  Deformation quantization of a classical mechanical system encoded by a Poisson manifold $(M,\{\cdot,\cdot\})$ means to find a star product on $\mathscr{A} = C^{\infty}(M)$ whose induced Poisson bracket equals the original one. Setting $\varepsilon = \frac{\I\hbar}{2}$, one obtains the canonical commutation relations \[ [f,g]_\star = \frac{\I\hbar}{2}\{f,g\} + O(\hbar^2)\] at first order in $\hbar$, where $[\cdot,\cdot]_\star$ is the commutator with respect to the star product.
\end{rem}

\subsection{Kontsevich's star product}
In \cite{K}, Kontsevich gave a general formula for a deformation quantization of any finite dimensional Poisson domain. 
\begin{thm}[Kontsevich]
\label{kontsevich}
Let $\alpha=\sum_{1\leq i<j\leq d}\alpha^{ij}(x)\partial_i\land\partial_j\in\Gamma\left(\bigwedge^2TU\right)$ be a Poisson structure\footnote{We write $\partial_i$ for $\frac{\partial}{\partial x^{i}}$.} on an open subset $U\subset\R^d$. Then for any two smooth functions $f,g\in C^\infty(U)$, there is an explicit star product, denoted by $\star_K$, given by the formula\footnote{This formula seems of course a bit strange without the precise derivation of the underlying objects. To get a full understanding of these objects we refer to \cite{K} or \cite{CKTB}.}
\[
f\star_Kg=\sum_{n\geq 0}\frac{\varepsilon^n}{n!}\sum_{\Gamma\in\mathcal{G}_n}w_K(\Gamma)B_{\Gamma,\alpha}(f,g),
\]
where $w_K(\Gamma)\in\R$ is the \textsf{Kontsevich weight} of an \textsf{admissible graph} $\Gamma\in \mathcal{G}_n$ defined as integration of some special angle $1$-forms on some configuration space of points on the upper half plane $\mathscr{H}=\{z\in\mathbb{C}\mid Im(z)>0\}$. Here $\mathcal{G}_n$ denotes a special set of graphs (satisfying some conditions), where the vertex set of such a graph is given by $n+2$ and the set of edges by $2n$ elements. The $B_{\Gamma,\alpha}(f,g)$ are bidifferential operators acting on $f$ and $g$, depending on the graph $\Gamma\in\mathcal{G}_n$ and the Poisson structure.
\end{thm}

\begin{defn}[Constant structure]
We say that a Poisson structure 
\[
\alpha(x)=\sum_{i<j}\alpha^{ij}(x)\partial_i\land\partial_j
\]
is \textsf{constant} if $\alpha^{ij}$ is a constant map for all $i,j$.
\end{defn}

\begin{cor}[Moyal product]
For $\alpha=\text{const}$, the star product of theorem \ref{kontsevich} coincides with the 
\textsf{Moyal product} \cite{Moy}
given by 
\begin{align*}
f\star_Mg&=fg+\varepsilon\sum_{i,j}\alpha^{ij}\partial_i(f)\partial_j(g)+\frac{\varepsilon^2}{2}\sum_{i,j,k,\ell}\alpha^{ij}\alpha^{k\ell}\partial_i\partial_k(f)\partial_j\partial_\ell(g)+O(\varepsilon^3)\\
&=\sum_{n\geq 0}\frac{\varepsilon^n}{n!}\sum_{i_1,...,i_n\atop j_1,...,j_n}\prod_{k=1}^n\alpha^{i_kj_k}\left(\prod_{k=1}^n\partial_{i_k}\right)(f)\times\left(\prod_{k=1}^n\partial_{j_k}\right)(g),
\end{align*}
where $\times$ denotes here the usual product.
\end{cor}

\begin{rem}
We can also write $\star_K\big|_{\alpha=\text{const}}=\star_M$. Moreover, one can write the Moyal product of two smooth functions $f,g\in C^\infty(\mathscr{P})$ as 
\[
f\star_M g(x)=\left.\ee^{\varepsilon \alpha^{ij}\frac{\partial}{\partial x^{i}}\frac{\partial}{\partial y^j}}f(x)g(y)\right |_{x=y}.
\]
One can then indeed easily check that associativity is satisfied.
\end{rem}

\section{The Poisson sigma model (PSM) and relation to Deformation Quantization}\label{s:PSM}

\subsection{Formulation of the model}
We want to recall the the Poisson sigma model and its connection to deformation quantization. Let $(\mathscr{P},\alpha)$ be a Poisson manifold with Poisson structure $\alpha\in \Gamma\left(\bigwedge^2T\mathscr{P}\right)$ and such that $\dim \mathscr{P}=d$. The data for the Poisson sigma model consists of a connected, oriented, smooth $2$-dimensional manifold $\Sigma$, called the \textsf{worldsheet}, and two fields. The fields are given by a  map $X\colon\Sigma\to \mathscr{P}$ and a $1$-form $\eta\in \Gamma(X^*(T^*\mathscr{P})\otimes T^*\Sigma)$. With this data we are able to define an action.
\begin{defn}[Poisson sigma model action]
The action functional for the Poisson sigma model is given by 
\[
S(X,\eta)=\int_\Sigma\eta_i(u)\land \dd X^{i}(u)+\frac{1}{2}\alpha^{ij}(X(u))\eta_i(u)\land\eta_j(u).
\]
In local coordinates we have $\eta_i(u)=\eta_{i\mu}\dd u^\mu$ for $u\in \Sigma$. 
\end{defn}

\subsection{Path integral formulation of Kontsevich's star product}
In \cite{CF1}, the first author and G. Felder have shown that Kontsevich's star product can be written as a path integral by using the Poisson sigma model action when the worldsheet is a disk. For this 
the boundary condition for $\eta$ is that for $u\in\partial \Sigma$ we get that $\eta_i(u)$ vanishes on vectors tangent to $\partial \Sigma$.

\begin{thm}
Let $(\mathscr{P},\alpha)$ be a Poisson manifold and let $D$ be the usual $2$-disk, i.e. $D=\{u\in\R^2\mid \|u\|\leq 1\}$, which we choose to be our worldsheet. Moreover, let $X$ and $\eta$ be the two fields of the Poisson sigma model described above. Then Kontsevich's star product of two smooth functions $f,g\in C^\infty(\mathscr{P})$ is given by the semiclassical expansion of the path integral
\[
f\star_Kg(x)=\int_{X(\infty)=x}f(X(1))g(X(0))\ee^{\frac{\I}{\hbar}S(X,\eta)}\mathscr{D}(X,\eta),
\]
where $0,1,\infty$ represent any three cyclically ordered\footnote{Cyclically ordered means that if we start from $0$ and move counterclockwise on the unit circle we will first meet $1$ and then $\infty$. One can also regard the unit circle here as a projective space of the real line where the point $\infty$ actually represents the identification of $a\to\infty$ and $a\to-\infty$ for $a\in\R$.} points on the unit circle (see figure \ref{cyc})

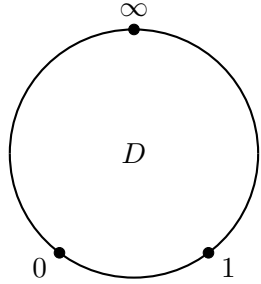
\begin{figure}[!ht]
\centering
\tikzset{
particle/.style={thick,draw=black},
particle2/.style={thick,draw=black, postaction={decorate},
    decoration={markings,mark=at position .9 with {\arrow[black]{triangle 45}}}},
gluon/.style={decorate, draw=black,
    decoration={coil,aspect=0}}
 }
\begin{tikzpicture}[x=0.04\textwidth, y=0.04\textwidth]
\node[](1) at (0,0){};
\node[](2) at (5,0){};
\node[](3) at (4,-2){};
\node[](4) at (4.4,-2.3){$1$};
\draw[fill=black] (3) circle (.07cm);
\node[](5) at (1,-2){};
\node[](6) at (0.6,-2.3){$0$};
\draw[fill=black] (5) circle (.07cm);
\node[](7) at (2.5,2.5){};
\node[](8) at (2.5,2.9){$\infty$};
\draw[fill=black] (7) circle (.07cm);
\node[](9) at (2.5,0){$D$};
\node[](2) at (5,0){};
\semiloop[particle]{1}{2}{0};
\semiloop[particle]{2}{1}{180};
\end{tikzpicture}
\caption{Cyclically ordered points on $S^1=\partial D$}
\label{cyc}
\end{figure}

\end{thm}

\section{Formal PSM in the BV-BFV formalism}\label{s:BVBFV}
Our goal is to give a perturbative description of the Poisson sigma model where we are allowed to vary the critical point around which we expand. Also gauge-fixing has to be performed. This has been discussed in the case of closed source manifolds in the literature (e.g. \cite{CF3,BCM}). We will recall these methods very briefly and then explain how to generalize them to the case with boundary, and apply the procedure to the   example $\mathscr{P}= \R^d$.
\subsection{Formal exponential maps and Grothendieck connections} 
\label{4.1}
Given a smooth manifold $M$, a generalized exponential map is a smooth map $\varphi$ from a neighbourhood $U \subset TM$ of the zero section to $M$, denoted $ (x,y) \mapsto \varphi_x(y)$, with $x \in M, y \in T_xM \cap U$ satisfying $\varphi_x(0) = x, \dd \varphi_x(0) = \id$. We identify two such maps if, for all $x \in M$, all their partial derivatives in the $y$ direction at the zero section coincide, and call such an equivalence class a \textsf{formal exponential map}. In local coordinates $\{x^i\}$ on the base around some point $x$ we can write such a formal exponential map as 
\begin{equation}
\label{formalexp.}
\varphi^{i}_x(y)=x^{i}+y^{i}+\frac{1}{2}\varphi^{i}_{x,jk}y^jy^k+\frac{1}{3!}\varphi_{x,jk\ell}^{i}y^jy^ky^\ell+\dotsm
\end{equation} 
As is explained in \cite{CF3,BCM}, such a formal exponential map induces a flat connection $D_{\textsf{G}}$, called \textsf{Grothendieck connection}, on $\Gamma(\widehat{S}T^*M)$, where $\widehat{S}$ denotes the completed symmetric algebra, with the property that $D_{\textsf{G}}\sigma = 0$ if and only if $\sigma(x) = T\varphi_x^*f$ is the Taylor expansion of the pullback of a function $f$ on $M$ at $y=0$. The connection describes how sections of this bundle behave under infinitesimal shifts on $M$, and its cohomology is concentrated in degree zero, where it coincides with the sections coming from global functions on $M$. In local coordinates $D_{\textsf{G}} = \dd x^i\frac{\partial}{\partial x^i} + \dd x^iR_i$ with 
$$ R_i = \left(\left(\frac{\partial \varphi_x}{\partial y}\right)^{-1}\right)^k_j\frac{\partial \varphi^j}{\partial x^i}\frac{\partial}{\partial y^k} =: Y^k_i\frac{\partial}{\partial y^k}.$$

\subsection{PSM in formal coordinates} 
\label{4.2}
The BV action for the Poisson sigma model with target $\mathscr{P}$ is 
\[
\mathcal{S}(\textsf{X},\boldsymbol\eta)=\int_\Sigma\left(\boldsymbol\eta_i(u)\land \dd \textsf{X}^{i}(u)+\frac{1}{2}\alpha^{ij}(\textsf{X}(u))\boldsymbol\eta_i(u)\land\boldsymbol\eta_j(u)\right).
\]
where
 $(\textsf{X}, \boldsymbol\eta) \in \mathrm{Map}(T[1]\Sigma,T^*[1]\mathscr{P})$ are the BV fields coming from the AKSZ formulation of the PSM (see \cite{AKSZ} or \cite{CF4}). Our goal is to expand around 
the $X$-component of the critical points of the kinetic term of the classical action,\footnote{\label{f:mod}Note that the whole
moduli space $\mathcal{M}$ of critical points of the kinetic term modulo symmetries is 
a vector bundle over $\mathscr{P}$ with fiber at $x\in\mathscr{P}$ given by $H^1(\Sigma)\otimes T^*_x\mathscr{P}$.
The fiber directions will be taken care of completely by the residual fields to be introduced below. Moreover, there is a canonical choice of background for $\eta$, namely, $\eta=0$. For these reasons we only take 
$\mathcal{M}_0$ as space of backgrounds.}
 i.e., a point of 
$$\mathcal{M}_0 = \{(X,\eta)| X(u) \equiv x \in \mathscr{P} \text{ const.} \}\cong\mathscr{P}.$$ 
The perturbative expansion around such a critical point only depends on a formal neighbourhood of it. Picking a formal exponential map $\varphi$ on $\mathscr{P}$ we can perform a change of coordinates in such a neighbourhood of this critical point:
$$ \mathsf{X} = \varphi_x(\hatX),\hspace{0.5cm} \boldsymbol{\eta} = \dd\varphi_x(\hatX)^{*,-1}\hateta $$ 
where $(\hatX,\hateta) \in \mathrm{Map}(T[1]\Sigma,T^*[1]T_x\mathscr{P})$. This can be interpreted as a formal exponential map on $\mathcal{M}_0 \subset \mathcal{F}$. In these coordinates the action reads (cf \cite{BCM}) 
$$ \mathcal{S}^{\varphi}_{\Sigma,x}[(\hatX, \hateta)] = \int_{\Sigma} \left(\hateta_i \wedge \dd\hatX^i + \frac{1}{2}(T\varphi_x^*\alpha)^{ij}(\hatX)\hateta_i\wedge\hateta_j\right). $$
We now want to gauge fix the model\footnote{One should do this before passing to formal coordinates but the result is the same, see \cite{CF3}.}. We first fix an embedding $H^\bullet(\Sigma)\hookrightarrow \Omega^\bullet(\Sigma)$, and think of $H^\bullet(\Sigma)$ as a subspace of $\Omega^\bullet(\Sigma)$ via this embedding. Now we define the space of residual fields\footnote{As anticipated in 
footnote~\ref{f:mod}, the ghost number zero component of the second summand takes care of the $\eta$-direction of the
moduli space of critical points. On the other hand, the ghost number zero component of the first summand in general
only sees a formal neighborhood of the $X$-component of the moduli space.} 
$$\mathcal{V}_{\Sigma,x} = H^{\bullet}(\Sigma) \otimes T_x\mathscr{P} \oplus H^{\bullet}(\Sigma) \otimes T^*_x\mathscr{P} $$ and a Lagrangian subspace $\mathcal{L}$ of a complement of $\mathcal{V}_{\Sigma,x} \subset \mathrm{Map}(T[1]\Sigma,T^*[1]T_x\mathscr{P})$. One can now formally define the partition function as 
$$\widehat{\psi}_{\Sigma,x} = \int_{\mathcal{L}}\ee^{\frac{\I}{\hbar}\mathcal{S}^{\varphi}_{\Sigma,x}}\in \mathrm{Dens}^{\frac{1}{2}}_{\mathrm{const}}(\mathcal{V}_{\Sigma,x})\otimes\widehat{S}^{\bullet}
\mathcal{V}^*_{\Sigma,x}[[\hbar]] =:\mathrm{Dens}^{\frac{1}{2}}_{\mathrm{formal}}(\mathcal{V}_{\Sigma,x}).$$ Under a change of gauge fixing it changes by a $\triangle$-exact term.
\subsection{Globalization}
\label{4.3}
We now look at the collection of the $\widehat{\psi}_{\Sigma,x}$ as a section of the bundle over $\mathcal{M}_0$ with fiber over $x$ given by  $\mathrm{Dens}^{\frac{1}{2}}_{\mathrm{formal}}(\mathcal{V}_{\Sigma,x}) $. If we vary $x$, it changes as (\cite{BCM})
$$ \dd_x\widehat{\psi}_{\Sigma,x} = \frac{\I}{\hbar}\int_{\mathcal{L}}\ee^{\frac{\I}{\hbar}\mathcal{S}^{\varphi}_{\Sigma,x}}(\mathcal{S}_R,\mathcal{S}^{\varphi}_{\Sigma,x}) $$
with $$\mathcal{S}_R =\int_\Sigma Y^j_i(\widehat{\textsf{X}})\widehat{\boldsymbol{\eta}}_j\land \dd_{\mathcal{M}_0}x^{i},$$
where $\dd_{\mathcal{M}_0}$ denotes the de Rham differential on $\mathcal{M}_0$. One can summarize the properties of formal geometry and the BV formalism by defining the \textsf{differential BV action} 
$$ \widetilde{\mathcal{S}}^{\varphi}_{\Sigma,x} = \mathcal{S}^{\varphi}_{\Sigma,x} + \mathcal{S}_R $$
which by construction satisfies the \textsf{differential classical master equation} (dCME)
\begin{equation} \dd_x\widetilde{\mathcal{S}}^{\varphi}_{\Sigma,x} + \frac{1}{2}(\widetilde{\mathcal{S}}^{\varphi}_{\Sigma,x},\widetilde{\mathcal{S}}^{\varphi}_{\Sigma,x}) = 0, \label{dCME} \end{equation} 
and the partition function 
$$ \widetilde{\psi}_{\Sigma,x} = \int_{\mathcal{L}}\ee^{\frac{\I}{\hbar}\widetilde{\mathcal{S}}^\varphi_{\Sigma,x}}$$ 
satisfying the \textsf{differential quantum master equation} (dQME) 
\begin{equation} \dd_x \widetilde{\psi}_{\Sigma,x} - \I\hbar\triangle\widetilde{\psi}_{\Sigma,x} = 0. \label{dQME} \end{equation}
We interpret equations \eqref{dCME},\eqref{dQME} as conditions on $\widetilde{\mathcal{S}}^\varphi_{\Sigma,x},\widetilde{\psi}_{\Sigma,x}$ to be pullbacks under a formal exponential map of global objects on BV manifolds, compare also Appendix F in \cite{CMR2}.

\subsection{Extension to the case with boundary}
\label{4.4}
In the presence of a boundary, equations \eqref{dCME} and \eqref{dQME} are no longer true. A good way to extend the BV-formalism to manifolds with boundary is the BV-BFV formalism discussed in \cite{CMR2,CMR1,CMW}. We will now propose\footnote{Similar computations were done in \cite{CS} for 1-dimensional gravity.} a generalization of it to the case of where the perturbative quantization is performed in families over the moduli space of classical solutions (i.e. suitable for globalization). This can be seen as an extension of the methods used in Appendix F in \cite{CMR2} to the case with boundary.
Namely, on a BV-BFV manifold $(\mathcal{F},\omega,Q,\mathcal{S},\pi)$ over an exact BFV manifold $(\mathcal{F}^{\partial},\omega^{\partial} = \delta \alpha^{\partial},Q^{\partial})$, one has the \textsf{modified classical master equation}
$$ \iota_Q\omega - \delta \mathcal{S} = \pi^*\alpha^\partial.$$
We expect the same equation to hold for the PSM on manifolds with boundary if we replace $\mathcal{S}$ by $\widetilde{\mathcal{S}}$ and $Q$ by $\widetilde{Q}$, where $\widetilde{Q} = Q + \widetilde{R}$ and $\widetilde{R}$ is the lift of $R$ to $\mathcal{F}$. Similarly, we expect the \textsf{modified quantum master equation} 
$$ (\hbar^2\triangle + \Omega)\widehat{\psi} = 0 $$ 
to hold in a family version over the moduli space of classical solutions. Namely, for any $x \in \mathscr{P}$ the PSM in formal coordinates around $x$ is a perturbation of two-dimensional abelian BF theory, so we can define the boundary BFV complex $(\mathcal{H}^{\mathcal{P}}_{\partial \Sigma,x},\Omega_{\de \Sigma,x})$ and space of states $$\widehat{\mathcal{H}}^{\mathcal{P}}_{\Sigma,x} = \mathcal{H}^{\mathcal{P}}_{\partial \Sigma,x} \otimes \mathrm{Dens}^{\frac{1}{2}}_{\mathrm{formal}}(\mathcal{V}_{\Sigma,x})$$ as in \cite{CMR2}, and define the bundle $\mathcal{H}^{tot} \to \mathcal{M}_0$ as the union of these fibers. On this bundle we define a connection\footnote{Note that here $\dd$ is the de Rham differential on $\mathcal{M}_0$.} 
\begin{equation}
\nabla^{\partial\Sigma}_{\textsf{G}} = \dd + \I\hbar\triangle + \frac{\I}{\hbar}\Omega
\end{equation}
which we call the \textsf{quantum Grothendieck connection}, and observe that it is flat, i.e. $(\nabla^{\partial\Sigma}_{\textsf{G}})^2\equiv0$. We then expect the partition function $\widetilde{\psi}_{\Sigma,x}$ to be a flat section with respect to this connection, i.e. to satisfy 
$$\left(\dd + \I\hbar\triangle + \frac{\I}{\hbar}\Omega\right)\widetilde{\psi}_{\Sigma,x} = 0. $$
We call this the \textsf{modified differential quantum master equation}. We will return to it in section \ref{mdQMEsection}.

\subsection{The special case $\mathscr{P}=\R^d$}
\label{4.6}
For the case of $\mathscr{P}=\R^d$ there is a simple formal exponential map given by (the equivalence class of) $\varphi_x(y) = x + y.$ We then get $R_i = -\frac{\partial}{\partial y^{i}}$, so that 
$$ D_{\textsf{G}} = \dd x^i\left(\frac{\partial}{\partial x^i} -\frac{\partial}{\partial y^i}\right). $$ 
The coordinates in a formal neighbourhood of $x \in \mathcal{M}_0$ are now defined by $\mathsf{X} = x + \hatX$ and $\boldsymbol\eta = \hateta$ with $(\hatX,\hateta) \in \mathrm{Map}(T[1]\Sigma,T^*[1]\R^d)$ and the action in these coordinates is $$ \mathcal{S}^{\varphi}_{\Sigma,x}[(\hatX, \hateta)] = \int_{\Sigma} \left(\hateta_i \wedge \dd\hatX^i + \frac{1}{2}\alpha^{ij}(x + \hatX)\hateta_i\wedge\hateta_j\right). $$
Here $\alpha(x + \hatX)$ is to be understood as the Taylor expansion of $\alpha$ in $\hatX$ around $x$. We have that 
$$\mathcal{S}_R = \int_{\Sigma} \widehat{\boldsymbol\eta}_i\land \dd_{\mathcal{M}_0} x^{i}$$ so that 
\begin{equation}
\label{action}
\widetilde{\mathcal{S}}^\varphi_{\Sigma,x} = \int_{\Sigma}\left(\widehat{\boldsymbol{\eta}}_i \land \dd_\Sigma \widehat{\textsf{X}}^{i}+\frac{1}{2}\alpha^{ij}(x+\widehat{\textsf{X}})\widehat{\boldsymbol\eta}_i
\land\widehat{\boldsymbol\eta}_j+\widehat{\boldsymbol\eta}_i\land \dd_{\mathcal{M}_0} x^{i}\right),
\end{equation}
where we want to emphasize that $\dd_\Sigma$ is the de Rham differential on $\Sigma$ and $\dd_{\mathcal{M}_0}$ the one on $\mathcal{M}_0\cong\mathscr{P}$. From now on we write for both differentials just $\dd$, where it should be clear from the context which one belongs to which space.
Letting $Q_0$ be the vector field of the BV theory given by $\mathcal{S}^{\varphi}_{\Sigma,x}$ and $Q=Q_0+\int_\Sigma \dd x^{i}\frac{\delta}{\delta{\widehat{\textsf{X}}^{i}}}$, the mdQME 
\begin{equation}
\label{mdCME}
\iota_Q\omega-\delta\widetilde{\mathcal{S}}^\varphi_{\Sigma,x}=\pi^*\alpha^\partial
\end{equation}
holds. Indeed, we have $\iota_Q\omega=\iota_{Q_0}\omega+\iota_{\int_\Sigma\dd x^{i}\frac{\delta}{\delta\widehat{\textsf{X}}}}\omega$ and since $\iota_{Q_0}\omega=\delta\mathcal{S}^{\varphi}_{\Sigma,x}+\pi^*\alpha^\partial$ we get 
\begin{multline*}
\iota_Q\omega-\delta\widetilde{\mathcal{S}}^\varphi_{\Sigma,x}=\iota_{Q_0}\omega+\iota_{\int_\Sigma\dd x^{i}\frac{\delta}{\delta\widehat{\textsf{X}}}}\omega-\delta\mathcal{S}^\varphi_{\Sigma,x} \\=\iota_{Q_0}\omega-\delta\mathcal{S}^\varphi_{\Sigma,x}+\iota_{\int_\Sigma\dd x^{i}\frac{\delta}{\delta\widehat{\textsf{X}}}}\omega-\delta\mathcal{S}_R=\pi^*\alpha^\partial+\underbrace{\int_\Sigma \dd x^{i}\land\delta\widehat{\boldsymbol{\eta}}_i}_{\delta\mathcal{S}_R}
-\delta\mathcal{S}_R=\pi^*\alpha^\partial,
\end{multline*}
Identifying tangent spaces to $\R^d$ at different points, the bundle $\mathcal{H}^{tot}$ becomes trivial (in general one needs an identification of tangent spaces at different points of $\mathscr{P}$ to do this). We can now apply the BV-BFV quantization procedure over every point $x\in \mathcal{M}_0$ by splitting the fields $\mathcal{F} =\mathcal{B}^{\mathcal{P}}_{\de \Sigma} \times \mathcal{V}_{\Sigma,x} \times \mathcal{Y}$ as in \cite{CMR2}
\begin{align*}
\widehat{\boldsymbol\eta}&=\E+\textsf{e}+\mathscr{E}\\
\widehat{\textsf{X}}&=\mathbb{X}+\textsf{x}+\mathscr{X},
\end{align*}
where $\E,\mathbb{X} \in \mathcal{B}^{\mathcal{P}}_{\de \Sigma}$ are the boundary fields, $\textsf{e},\textsf{x} \in \mathcal{V}_{\Sigma,x}$ are the residual fields and $\mathscr{E},\mathscr{X}\in \mathcal{Y}$ are the fluctuation fields. We then proceed to pick a gauge-fixing Lagrangian $\mathcal{L} \subset  \mathrm{Map}(T[1]\Sigma,T^*[1]\R^d)\ni (\hatX,\hateta)$. Finally we are interested in the state 
$$\widehat{\psi}_{\Sigma,x}(\E,\textsf{e},\mathbb{X},\textsf{x}) = \int_{(\mathscr{E},\mathscr{X})\in\calL}\ee^{\frac{\I}{\hbar}\widetilde{\mathcal{S}}^{\varphi}_{\Sigma,x}\left[(\widehat{\textsf{X}},\widehat{\boldsymbol\eta})\right]}.$$

\begin{rem}
Note that this entire section holds for a general smooth Poisson structure on $\R^d$, as long as one uses the trivial formal exponential map. 
\end{rem}

\section{The Relational Symplectic Groupoid}\label{s:RSG}

\subsection{Short description of the RSG}
Symplectic groupoids are an important concept in Poisson and symplectic geometry (\cite{W}).
A groupoid is a small category whose morphisms are invertible. 
We denote a groupoid
by $G\rightrightarrows M$, where $M$ is the set of objects and $G$ the set of morphisms.
A Lie groupoid is, roughly speaking, a groupoid where 
$M$ and $G$ are smooth manifolds and all structure maps are smooth. Finally, a symplectic groupoid is a Lie groupoid with a symplectic form $\omega\in\Omega^2(G)$ such that the graph of the multiplication  is a Lagrangian submanifold of $(G,\omega)\times(G,\omega)\times(G,-\omega)$. The manifold of objects $M$ has an induced Poisson
structure uniquely determined by requiring that the source map $G\to M$ is Poisson. A Poisson manifold $M$ that arises this way is called integrable. Not every Poisson manifold is integrable. 

The reduced phase space of the PSM on a boundary interval
with target an integrable Poisson manifold $\mathscr{P}$ is the source simply connected symplectic groupoid of $\mathscr{P}$ (\cite{CF5}). In general, the reduced phase space is a topological groupoid arising by singular symplectic reduction. In (\cite{C,CC1,CC2}) it was however shown that the space of 
classical boundary fields
always
has an interesting structure called \textsf{relational symplectic groupoid} (RSG). 
An RSG is, roughly speaking, a groupoid in the ``extended category'' of symplectic manifolds where morphisms
are canonical relations. Recall that a canonical relation from  $(M_1,\omega_1)$ to
$(M_2,\omega_2)$ is an immersed Lagrangian submanifold of 
$(M_1,\omega_1)\times(M_2,-\omega_2)$.
The main structure of an RSG $(\mathcal{G},\omega)$ is then given by an immersed Lagrangian submanifold 
$\mathcal{L}_1$ of $(\mathcal{G},\omega)$, which plays the role of unity, and by
an immersed Lagrangian submanifold $\mathcal{L}_3$ of
 $(\mathcal{G},\omega)\times(\mathcal{G},\omega)\times(\mathcal{G},-\omega)$,
 which plays the role
 of associative multiplication. (In addition, there is also an antisymplectomorphism $\mathcal{I}$ of $\mathcal{G}$ that plays the role of the inversion map.)
In case $M$ is integrable, it was also shown that the RSG $\mathcal{G}$ is equivalent, as an RSG, to the the symplectic groupoid $G$.



%

\subsection{Description of the canonical relations}
For a boundary interval $I$ and a target Poisson manifold $\mathscr{P}$, the space of classical boundary fields $\mathcal{G}$ of the PSM
is the space of bundle maps $TI\to T^*\mathscr{P}$,
 which can be identified with (a version of) the cotangent bundle of the path space $\text{Map}(I,\mathscr{P})$ with its canonical symplectic 
 form $\omega$.\footnote{If we take smooth maps, then $(\mathcal{G},\omega)$ is a weak-symplectic Fr\'echet manifold.}
 For $n \geq 1 $, we denote by $L_n$ the disk whose boundary $S^{1}$ splits into $2n$ closed intervals $I$ intersecting only at the end points and with the boundary condition $\widehat{\boldsymbol{\eta}}=0$ on alternating intervals. The remaining $n$ intervals are free, so the space of classical boundary fields
 associated to $L_n$
  is $F_{L_n}^\partial:=(\mathcal{G},\omega)^{n}$ 
 if all the intervals are given the induced orientation.   We select however one of the intervals as ``outgoing'', meaning that we reverse its orientation: the corresponding space of classical boundary fields is then $\bar F_{L_n}^\partial:=(\mathcal{G},\omega)^{n-1}\times(\mathcal{G},-\omega)$. 
  (Note that $\bar F_{L_n}^\partial$
 can also be obtained from $F_{L_n}^\partial$ by applying to the component corresponding to the outgoing interval the antisymplectomorphism
 $\mathcal{I}$ given by pulling back the fields by $\phi\colon[a,b]\to[a,b]$, $t\mapsto a+b-t$.)
 Finally, 
 $\mathcal{L}_n\subset\bar F_{L_n}^\partial$
 is defined as the space of restrictions to the boundary of solutions to the Euler--Lagrange equation on $L_n$.
 The main result of 
 (\cite{C,CC2}) is that each $\mathcal{L}_n$ is a canonical relation and that the two ways of composing two $\mathcal{L}_3$s are identical to each other (and to
 $\mathcal{L}_4$).
 
 Finally, recall that a Lagrangian submanifold (a canonical relation) is the classical version of a state (an operator). 
 Roughly speaking, in this note
 we will construct the states $\widehat{\psi}_{L_n}$
corresponding to the quantization of $\mathcal{L}_n$ in the case of a constant Poisson structure on the target and
 we will show that, in the $\nabla^{\partial\Sigma}_{\textsf{G}}$-cohomology, the two ways of composing two  $\widehat{\psi}_{L_3}$s are identical to each other (and to
 $\widehat{\psi}_{L_4}$).

%

\subsection{Deformation quantization of the RSG}
\label{4.5}
In \cite{CMR2} a procedure for the deformation quantization of the relational symplectic groupoid 
in the case when the target Poisson manifold is $\mathscr{P}=\mathbb{R}^d$
was introduced. Let us repeat the main points. 
The space of BV boundary fields corresponding to $L_n$
is  $\F_{L_n}^\partial=(\F_I^\partial)^n$ with 
\[
\F_I^\partial=\Omega^\bullet(I)\otimes \R^d\oplus\Omega^\bullet_0(I)\otimes(\R^d)^*[1],
\]
with $\Omega_0^\bullet(I)$ denoting the subcomplex of forms whose restriction to the end points is zero. Choose a polarization $\calP$ of $\F^{\partial}_I$, and denote $\overline{\calP}$ the opposite polarization.

Denote by $\mathcal{H}$ the vector space which quantizes $\F^\partial_I$ in this polarization. Compute the state $m_x$ associated to $L_3 $ with polarization $\calP \times \overline{\calP} \times \overline{\calP}$ perturbing around a constant solution $X=x$. We can see it as a linear map $\mathcal{H}\otimes\mathcal{H}\to\mathcal{H}$ (see figure \ref{L3_mode}). Next we observe that there are two inequivalent ways to cut $L_4$ into gluings of two $L_3$s. From this we see that $m_x$ defines an associative structure in the $(\hbar^2\triangle+\Omega)$-cohomology for $L_4$. This provides a way of defining the deformation quantization of the relational symplectic groupoid (see figure \ref{assL_3}).
To compare the result with the deformation quantization of the Poisson manifold $\mathscr{P}$, we have to  ``glue caps'', i.e. consider also $L_1$ (see e.g. figure \ref{gluing_moyal}). We view the state $\sigma_x$ associated to it as a linear map $\mathcal{H}\to\mathbb{C}[[\varepsilon]]$, with $\varepsilon=\frac{\I\hbar}{2}$. If $f$ is a function on $\mathscr{P}$, we may also take the expectation value of the observable $f(X(u_0))$, where $u_0$ is a point in the interior of the interval with the boundary condition. We denote the result by $\tau_xf$. We may view $\tau_x$ as a linear map $C^\infty(\mathscr{P})\otimes\mathbb{C}[[\varepsilon]]\to\mathcal{H}$. Kontsevich's star product is  obtained by the composition
\[
(f\star_Kg)(x)=\sigma_x(m_x(\tau_xf\otimes\tau_xg)).
\]

\section{The states for the RSG}

For a constant Poisson structure we will now show how the Moyal product can be obtained out of the relational symplectic groupoid by using the BV-BFV quantization formalism.

\subsection{Spaces of residual fields}\label{Residual fields}

We will begin by briefly discussing spaces of residual fields on $L_n$. For this we need the following simple fact. If we denote by $D$ the unit disk and by $I\subsetneq \partial D$ a closed interval, then $H^{\bullet}(D,I) = H^{\bullet}(D,\de D \setminus I) = \{0\}$, since the disk contracts onto $I$\footnote{We thank the referee for pointing out this simple argument.}. It follows that as soon as we have one interval on $\de D$ with boundary condition $\mathsf{x} = 0$ the spaces of residual fields vanish. This happens if we choose the $\frac{\delta}{\delta\E}$-polarization on one of the intervals\footnote{Recall that $\mathbb{X}$ and $\mathbb{E}$ denote the $\widehat{\textsf{X}}$ and $\widehat{\boldsymbol{\eta}}$ components of boundary fields respectively.}. If we choose the $\frac{\delta}{\delta\mathbb{X}}$-polarization on every interval, the space of residual fields will be 
$$\mathcal{V}_D = H^{\bullet}(D) \otimes \R^d \oplus H^{\bullet}(D,\de D) \otimes \R^d[1] \cong T^*[-1]\R^d. $$
\subsection{The state for $L_3$}
 
\begin{figure}[!ht]
\centering
\tikzset{
particle/.style={thick,draw=black},
particle2/.style={thick,draw=blue},
avector/.style={thick,draw=black, postaction={decorate},
    decoration={markings,mark=at position 1 with {\arrow[black]{triangle 45}}}},
gluon/.style={decorate, draw=black,
    decoration={coil,aspect=0}}
 }
\begin{tikzpicture}[x=0.05\textwidth, y=0.05\textwidth]
\draw[particle] (0,0)--(2,5) node[above]{};
\draw[particle2] (2,5)--(4,5) 
node[right]{};
\draw[particle2] (0,0)--(2,0) node[right]{};
\node[](u1) at (2,0){};
\node[](u2) at (4,0){};
\draw[particle2] (4,0)--(6,0)
node[right]{};
\draw[particle] (6,0)--(4,5) node[right]{};
\node[](2) at (0.7,0){};
\node[](3) at (1.4,0){};
\node[](eta1) at (0,2.5){$\widehat{\boldsymbol{\eta}}=0$};
\node[](eta2) at (6,2.5){$\widehat{\boldsymbol{\eta}}=0$};
\node[](eta3) at (3,0){$\widehat{\boldsymbol{\eta}}=0$};
\node[](e1) at (1,-0.5){$\partial_2^{(1)}M$};
\node[](eta1) at (1,0.5){$\E^{(1)}$};
\node[](e2) at (5,-0.5){$\partial_2^{(2)}M$};
\node[](e2) at (5,0.5){$\E^{(2)}$};
\node[](e3) at (3,5.5){$\partial_1M$};
\node[](x) at (3,4.5){$\mathbb{X}$};
\semiloop[particle]{u1}{u2}{0};
\end{tikzpicture}
\caption{The polarization for $L_3$}
\label{L3_mode}
\end{figure}
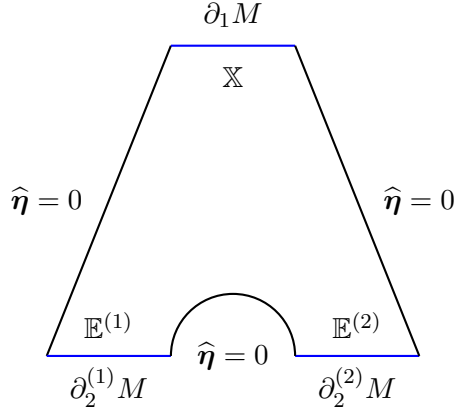

Let us start with $M=L_3$. Consider the polarization to be given as in figure \ref{L3_mode} such that we have the boundary field $\widehat{\textsf{X}}\big|_{\partial_1M}=\mathbb{X}$ on $\partial_1M$ and the boundary fields $\widehat{\boldsymbol{\eta}}\big|_{\partial_2^{(1)}M}=\E^{(1)}$ and $\widehat{\boldsymbol{\eta}}\big|_{\partial_2^{(2)}M}=\E^{(2)}$ on $\partial_2^{(1)}M$ and $\partial_{2}^{(2)}M$ respectively, where $\partial_2M:=\partial_2^{(1)}M\sqcup\partial_2^{(2)}M$. This actually means that we choose the $\frac{\delta}{\delta\mathbb{X}}$-polarization on $\partial_2M$ and the $\frac{\delta}{\delta\E}$-polarization on $\partial_1M$. The boundary condition for $\widehat{\boldsymbol{\eta}}$ is such that it is zero on the black boundary components.
 
\begin{figure}[!ht]
\centering
\subfigure[Free term for $\partial_2^{(1)}M$\label{subfigfreede1}]{
\tikzset{
particle/.style={thick,draw=black},
particle2/.style={thick,draw=blue},
avector/.style={thick,draw=black, postaction={decorate},
    decoration={markings,mark=at position 1 with {\arrow[black]{triangle 45}}}},
avector2/.style={thick,draw=red, postaction={decorate},
    decoration={markings,mark=at position 1 with {\arrow[red]{triangle 45}}}},
gluon/.style={decorate, draw=black,
    decoration={coil,aspect=0}}
 }
\begin{tikzpicture}[x=0.04\textwidth, y=0.04\textwidth]
\draw[particle] (0,0)--(2,5) node[above]{};
\draw[particle2] (2,5)--(4,5) node[right]{};
\draw[particle2] (0,0)--(2,0) node[right]{};
\node[](u1) at (2,0){};
\node[](u2) at (4,0){};
\draw[particle2] (4,0)--(6,0) node[right]{};
\draw[particle] (6,0)--(4,5) node[right]{};
\node[](1) at (3,5){};
\node[](2) at (1,0){};
\node[](e1) at (1,-0.5){$\E^{(1)}$};
\node[](e2) at (5,-0.5){$\E^{(2)}$};
\node[](e3) at (3,5.5){$\mathbb{X}$};
\draw[avector2] (3,5)--(1,0) node[right]{};
\semiloop[particle]{u1}{u2}{0};
\end{tikzpicture}
}
\quad
\subfigure[Free term for $\partial_2^{(2)}M$\label{subfigfreede2}]{
\tikzset{
particle/.style={thick,draw=black},
particle2/.style={thick,draw=blue},
avector/.style={thick,draw=black, postaction={decorate},
    decoration={markings,mark=at position 1 with {\arrow[black]{triangle 45}}}},
avector2/.style={thick,draw=red, postaction={decorate},
    decoration={markings,mark=at position 1 with {\arrow[red]{triangle 45}}}},
gluon/.style={decorate, draw=black,
    decoration={coil,aspect=0}}
 }
\begin{tikzpicture}[x=0.04\textwidth, y=0.04\textwidth]
\draw[particle] (0,0)--(2,5) node[above]{};
\draw[particle2] (2,5)--(4,5) node[right]{};
\draw[particle2] (0,0)--(2,0) node[right]{};
\node[](u1) at (2,0){};
\node[](u2) at (4,0){};
\draw[particle2] (4,0)--(6,0) node[right]{};
\draw[particle] (6,0)--(4,5) node[right]{};
\node[](1) at (3,5){};
\node[](2) at (5,0){};
\node[](e1) at (1,-0.5){$\E^{(1)}$};
\node[](e2) at (5,-0.5){$\E^{(2)}$};
\node[](e3) at (3,5.5){$\mathbb{X}$};
\draw[avector2] (3,5)--(5,0) node[right]{};
\semiloop[particle]{u1}{u2}{0};
\end{tikzpicture}
}
\quad
\subfigure[Interaction term for $\partial_2^{(1)}M$\label{subfigintde1}]{
\tikzset{
particle/.style={thick,draw=black},
particle2/.style={thick,draw=blue},
avector/.style={thick,draw=black, postaction={decorate},
    decoration={markings,mark=at position 1 with {\arrow[black]{triangle 45}}}},
avector2/.style={thick,draw=red, postaction={decorate},
    decoration={markings,mark=at position 1 with {\arrow[black]{triangle 45}}}},
gluon/.style={decorate, draw=black,
    decoration={coil,aspect=0}}
 }
\begin{tikzpicture}[x=0.04\textwidth, y=0.04\textwidth]
\draw[particle] (0,0)--(2,5) node[above]{};
\draw[particle2] (2,5)--(4,5) node[right]{};
\draw[particle2] (0,0)--(2,0) node[right]{};
\node[](u1) at (2,0){};
\node[](u2) at (4,0){};
\draw[particle2] (4,0)--(6,0) node[right]{};
\draw[particle] (6,0)--(4,5) node[right]{};
\node[](1) at (3,3){};
\draw[fill=black] (1) circle (0.05cm);
\node[](2) at (0.7,0){};
\node[](3) at (1.4,0){};
\node[](e1) at (1,-0.5){$\E^{(1)}$};
\node[](e2) at (5,-0.5){$\E^{(2)}$};
\node[](e3) at (3,5.5){$\mathbb{X}$};
\draw[avector] (3,3)--(0.5,0);
\draw[avector] (3,3)--(1.5,0);
\semiloop[particle]{u1}{u2}{0};
\end{tikzpicture}}
\quad
\subfigure[Interaction term for $\partial_2^{(2)}M$\label{subfigintde2}]{
\tikzset{
particle/.style={thick,draw=black},
particle2/.style={thick,draw=blue},
avector/.style={thick,draw=black, postaction={decorate},
    decoration={markings,mark=at position 1 with {\arrow[black]{triangle 45}}}},
gluon/.style={decorate, draw=black,
    decoration={coil,aspect=0}}
 }
\begin{tikzpicture}[x=0.04\textwidth, y=0.04\textwidth]
\draw[particle] (0,0)--(2,5) node[above]{};
\draw[particle2] (2,5)--(4,5) node[right]{};
\draw[particle2] (0,0)--(2,0) node[right]{};
\node[](u1) at (2,0){};
\node[](u2) at (4,0){};
\draw[particle2] (4,0)--(6,0) node[right]{};
\draw[particle] (6,0)--(4,5) node[right]{};
\node[](1) at (3,3){};
\draw[fill=black] (1) circle (0.05cm);
\node[](2) at (5.5,0){};
\node[](3) at (4.5,0){};
\node[](e1) at (1,-0.5){$\E^{(1)}$};
\node[](e2) at (5,-0.5){$\E^{(2)}$};
\node[](e3) at (3,5.5){$\mathbb{X}$};
\draw[avector] (3,3)--(5.5,0) node[right]{};
\draw[avector] (3,3)--(4.5,0) node[right]{};
\semiloop[particle]{u1}{u2}{0};
\end{tikzpicture}
}
\subfigure[Interaction term for $\partial_2^{(1)}M$ and $\partial_2^{(2)}M$\label{subfigintde12}]{
\tikzset{
particle/.style={thick,draw=black},
particle2/.style={thick,draw=blue},
avector/.style={thick,draw=black, postaction={decorate},
    decoration={markings,mark=at position 1 with {\arrow[black]{triangle 45}}}},
gluon/.style={decorate, draw=black,
    decoration={coil,aspect=0}}
 }
\begin{tikzpicture}[x=0.04\textwidth, y=0.04\textwidth]
\draw[particle] (0,0)--(2,5) node[above]{};
\draw[particle2] (2,5)--(4,5) node[right]{};
\draw[particle2] (0,0)--(2,0) node[right]{};
\node[](u1) at (2,0){};
\node[](u2) at (4,0){};
\draw[particle2] (4,0)--(6,0) node[right]{};
\draw[particle] (6,0)--(4,5) node[right]{};
\node[](1) at (3,3){};
\draw[fill=black] (1) circle (0.05cm);
\node[](2) at (1,0){};
\node[](2) at (5,0){};
\node[](e1) at (1,-0.5){$\E^{(1)}$};
\node[](e2) at (5,-0.5){$\E^{(2)}$};
\node[](e3) at (3,5.5){$\mathbb{X}$};
\draw[avector] (3,3)--(5,0) node[right]{};
\draw[avector] (3,3)--(1,0) node[right]{};
\semiloop[particle]{u1}{u2}{0};
\end{tikzpicture}
}
\caption{The diagrams for $L_3$ which need to be computed. The diagrams (a) and (b) are the free terms of the action and (c),(d) and (e) are interaction terms.}
\label{diagrams}
\end{figure}
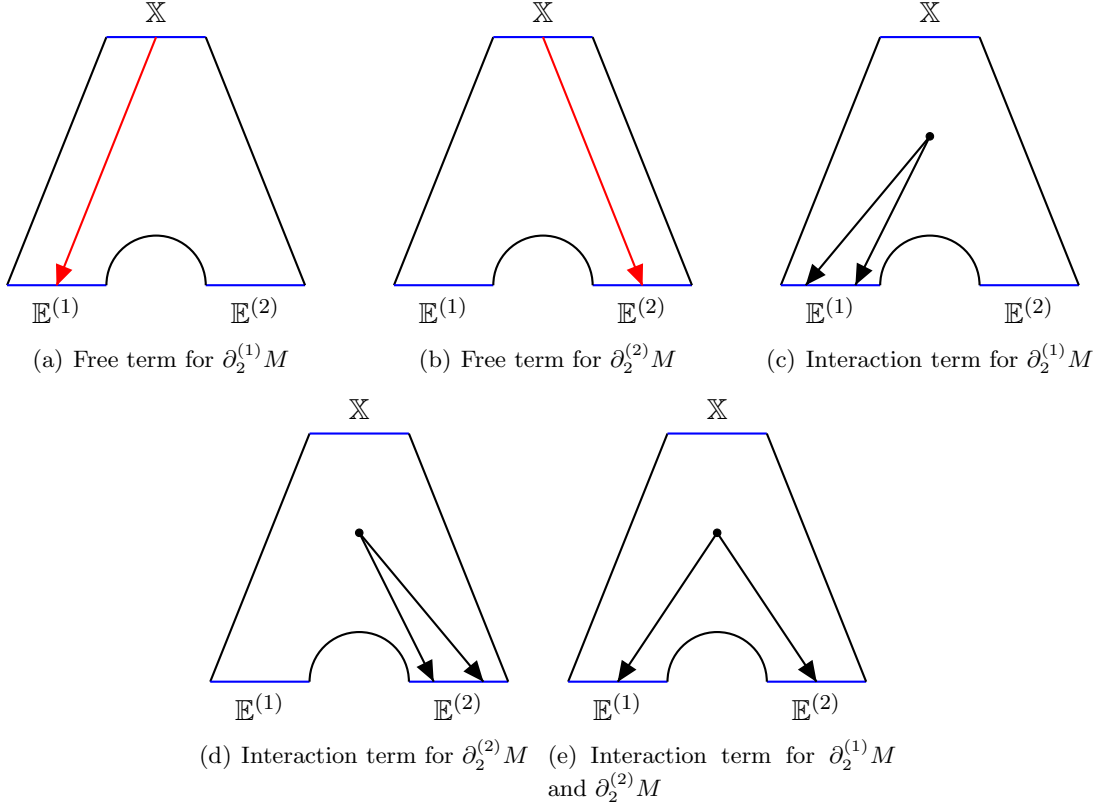

 As discussed above in this case the relevant cohomology vanishes and therefore there are no residual fields. Since $\alpha$ is constant the only Feynman diagrams that we have to compute are given as in figure \ref{diagrams}. Let us denote by $\zeta$ a propagator on the disk with the above boundary conditions and let us go with the convention that indices $ij$ on the propagator $\zeta$  represent the points $u_{i}$ and $u_j$ on the disk where $\zeta$ is evaluated as a two point function. Moreover, let the index $1$ always represent a point in the bulk and $2,3$ the points on the respective boundary component of $\partial_2M$. Moreover, let the index $0$ always represent a point on $\partial_1M$. Following \cite{CMR2} we compute the effective action. We start with the diagrams \ref{subfigfreede1} and \ref{subfigfreede2}. These are the free parts of the action 
\[
\mathcal{S}^{\textnormal{eff}}_{\partial^{(k)} M}=-\int_{\partial_2^{(k)}M\times \partial_1M}\pi_1^*\E_i\land\zeta_{0\nu_k}\land\pi_2^*\mathbb{X}_i,
\]
with $u_{\nu_k}\in \partial_2^{(k)}M$ for $k\in\{1,2\}$. Here $\pi_1$ and $\pi_2$ represent the projection onto the first and second component respectively. The perturbation term in \ref{subfigintde1} is obtained by the integral 
\[
\mathcal{S}^{\textnormal{pert,eff}}_{\partial_2^{(1)} M}=\frac{1}{2}\alpha^{ij}\int_{M\times C_2( \partial_2^{(1)}M)}\zeta_{12}\land\zeta_{13}\land\pi_{2,1}^*\E_{i}\land\pi_{2,2}^*\E_{j},
\]
where $\pi_{2,1}$ is the projection onto the first component of the configuration space $C_2(\partial_2^{(1)}M)$, $\pi_{2,2}$ the projection onto the second component of $C_2(\partial_2^{(1)}M)$. Integration over the bulk vertex gives then 
\[
\mathcal{S}^{\textnormal{pert,eff}}_{\partial_2^{(1)} M}=\frac{1}{2}\alpha^{ij}\int_{C_2(\partial_2^{(1)}M)}\pi_1^*\E_i\land\widehat{\zeta}^\partial_{23}\land\pi_2^*\E_j
\]
where $\widehat{\zeta}^\partial\in\Omega^0(\partial_2M)$ is given by $\widehat{\zeta}^\partial(u_2,u_3)=\int_{M}\zeta_{12}\land \zeta_{13}$. Since the propagator is a 1-form, this function is skew-symmetric. It vanishes if one of its arguments is on the boundary of the interval (since the propagator does). By Stokes theorem we have  $\dd\widehat{\zeta}^\partial(u_2,u_3)=\int_{\partial_1M}\zeta_{02}\land\zeta_{03}$ (since the propagator is closed) which implies, together with the above, that 
$$ \lim_{s \to t^{-}} \widehat{\zeta}^{\partial}(s,t) - \lim_{s \to t^{+}} \widehat{\zeta}^{\partial} (s,t) = 1.  $$  
These properties are important for the proof of the mdQME (see \ref{appendA}, subsection \ref{comp_L3}). Similarly for \ref{subfigintde2} we get the perturbation term 
\[
\mathcal{S}^{\textnormal{pert,eff}}_{\partial_2^{(2)} M}=\frac{1}{2}\alpha^{ij}\int_{C_2(\partial_2^{(2)}M)}\pi_1^*\E_i\land\widehat{\zeta}^\partial_{23}\land\pi_2^*\E_j.
\]
The perturbation term for \ref{subfigintde12} is given by 
\[
\mathcal{S}^{\textnormal{pert,eff}}_{\partial_2^{(1)}M\sqcup\partial_2^{(2)}M}=\frac{1}{2}\alpha^{ij}\int_{\partial_2^{(1)}M\times \partial_2^{(2)}M}\pi_1^*\E_i\land\widehat{\zeta}^\partial_{23}\land\pi_2^*\E_j.
\]
The state for $L_3$ is then given by 
\[
\widehat{\psi}_{L_3}=T_{L_3}\ee^{\frac{\I}{\hbar}\left(\mathcal{S}^{\textnormal{eff}}_{\partial^{(1)} M}+\mathcal{S}^{\textnormal{eff}}_{\partial^{(2)} M}+\mathcal{S}^{\textnormal{pert,eff}}_{\partial_2^{(1)} M}+\mathcal{S}^{\textnormal{pert,eff}}_{\partial_2^{(2)} M}+\mathcal{S}^{\textnormal{pert,eff}}_{\partial_2^{(1)}M\sqcup\partial_2^{(2)}M}\right)}=T_{L_3}\ee^{\frac{\I}{\hbar}\mathcal{S}^{\textnormal{eff}}_{\partial M}}.
\]
Here $$T_{M} = \int_{\mathcal{L}} \ee^{\frac{\I}{\hbar}\left(\int_{M}\mathscr{E}\wedge\dd \mathscr{X}\right)} \in \mathbb{C} \otimes \mathrm{Dens}^{\frac{1}{2}}_{\mathrm{const}}(\mathcal{V}_M)/\{\pm 1\}$$ is a regularized Gaussian functional integral related to the torsion of $M$, described in \cite{CMR2}. Since the disk is simply connected and the relative cohomology for these boundary conditions is trivial, we have that $T_{L_3}=1$.

\subsection{The state for $L_1$}
 
\begin{figure}[!ht]
\centering
\subfigure[Free term appearing due to \newline cohomology with the $\frac{\delta}{\delta\mathbb{X}}$-polarization.\label{subfigfreexEddX}]{
\tikzset{
particle/.style={thick,draw=black},
particle2/.style={thick,draw=blue},
avector/.style={thick,draw=black, postaction={decorate},
    decoration={markings,mark=at position 1 with {\arrow[black]{triangle 45}}}},
gluon/.style={decorate, draw=black,
    decoration={coil,aspect=0}}
 }
\begin{tikzpicture}[x=0.05\textwidth, y=0.05\textwidth]
\draw[particle2] (0,-2.5)--(5,-2.5) node[above]{};
\node[](u1) at (0,-2.5){};
\node[](u2) at (5,-2.5){};
\node[label=below:$\mathsf{x}\land\E$](eta1) at (2.5,-2.5){};
\node[](eta2) at (2.5,0.5){$\widehat{\boldsymbol{\eta}}=0$};
\draw[fill=black] (eta1) circle (0.05cm);
\semiloop[particle]{u1}{u2}{0};
\node(n) at (-1,-2.5){$\partial_2M$};
\end{tikzpicture}
}
\subfigure[Free term appearing due to \newline cohomology with the $\frac{\delta}{\delta\mathbb{X}}$-polarization.\label{subfigfreemuddX}]{
\tikzset{
particle/.style={thick,draw=black},
particle2/.style={thick,draw=blue},
avector/.style={thick,draw=black, postaction={decorate},
    decoration={markings,mark=at position 1 with {\arrow[black]{triangle 45}}}},
gluon/.style={decorate, draw=black,
    decoration={coil,aspect=0}}
 }
\begin{tikzpicture}[x=0.05\textwidth, y=0.05\textwidth]
\draw[particle2] (0,-2.5)--(5,-2.5) node[above]{};
\node[](u1) at (0,-2.5){};
\node[](u2) at (5,-2.5){};
\node[label=above:$\mathsf{e} \land\dd x$](2) at (2.5,-1){};
\node[label=below:$\E$](eta1) at (2.5,-2.5){};
\draw[fill=black] (2) circle (0.05cm);
\node[](eta2) at (2.5,0.5){$\widehat{\boldsymbol{\eta}}=0$};
\semiloop[particle]{u1}{u2}{0};
\node(n) at (-1,-2.5){$\partial_2M$};
\end{tikzpicture}
}
\subfigure[Usual interaction term \newline with the $\frac{\delta}{\delta\mathbb{X}}$-polarization.\label{subfigintddX}]{
\tikzset{
particle/.style={thick,draw=black},
particle2/.style={thick,draw=blue},
avector/.style={thick,draw=black, postaction={decorate},
    decoration={markings,mark=at position 1 with {\arrow[black]{triangle 45}}}},
gluon/.style={decorate, draw=black,
    decoration={coil,aspect=0}}
 }
\begin{tikzpicture}[x=0.05\textwidth, y=0.05\textwidth]
\draw[particle2] (0,-2.5)--(5,-2.5) node[above]{};
\node[](u1) at (0,-2.5){};
\node[](u2) at (5,-2.5){};
\node[](2) at (2.5,-1){};
\draw[fill=black] (2) circle (0.05cm);
\draw[avector] (2.5,-1)--(1,-2.5) node[above]{};
\draw[avector] (2.5,-1)--(4,-2.5) node[above]{};
\node[](eta1) at (2.5,-3){$\E$};
\node[](eta2) at (2.5,0.5){$\widehat{\boldsymbol{\eta}}=0$};
\semiloop[particle]{u1}{u2}{0};
\node(n) at (-1,-2.5){$\partial_2M$};
\end{tikzpicture}
}
\subfigure[Interaction term appearing due to cohomology with the $\frac{\delta}{\delta\mathbb{X}}$-polarization.\label{subfigintmuddX}]{
\tikzset{
particle/.style={thick,draw=black},
particle2/.style={thick,draw=blue},
avector/.style={thick,draw=black, postaction={decorate},
    decoration={markings,mark=at position 1 with {\arrow[black]{triangle 45}}}},
gluon/.style={decorate, draw=black,
    decoration={coil,aspect=0}}
 }
\begin{tikzpicture}[x=0.05\textwidth, y=0.05\textwidth]
\draw[particle2] (0,-2.5)--(5,-2.5) node[above]{};
\node[](u1) at (0,-2.5){};
\node[](u2) at (5,-2.5){};
\node[label=above:$\mu$](2) at (2.5,-1){};
\draw[fill=black] (2) circle (0.05cm);
\draw[avector] (2.5,-1)--(2.5,-2.5) node[above]{};
\node[](eta1) at (2.5,-3){$\E$};
\node[](eta2) at (2.5,0.5){$\widehat{\boldsymbol{\eta}}=0$};
\semiloop[particle]{u1}{u2}{0};
\node(n) at (-1,-2.5){$\partial_2M$};
\end{tikzpicture}
}
\subfigure[No diagram for the $\frac{\delta}{\delta\mathbb{E}}$-polarization.\label{subfigddEtrivial}]{
\tikzset{
particle/.style={thick,draw=black},
particle2/.style={thick,draw=blue},
avector/.style={thick,draw=black, postaction={decorate},
    decoration={markings,mark=at position 1 with {\arrow[black]{triangle 45}}}},
gluon/.style={decorate, draw=black,
    decoration={coil,aspect=0}}
 }
\begin{tikzpicture}[x=0.05\textwidth, y=0.05\textwidth]
\draw[particle2] (0,0)--(5,0) node[above]{};
\node[](u1) at (0,0){};
\node[](u2) at (5,0){};
\node[](2) at (2.5,-2){};
\node[](eta1) at (2.5,0.4){$\mathbb{X}$};
\node[](eta2) at (2.5,-3){$\widehat{\boldsymbol{\eta}}=0$};
\semiloop[particle]{u2}{u1}{180};
\node(n) at (6,0){$\partial_1 M$};
\end{tikzpicture}}
\caption{The diagrams for $L_1$.}
\label{L1}
\end{figure}
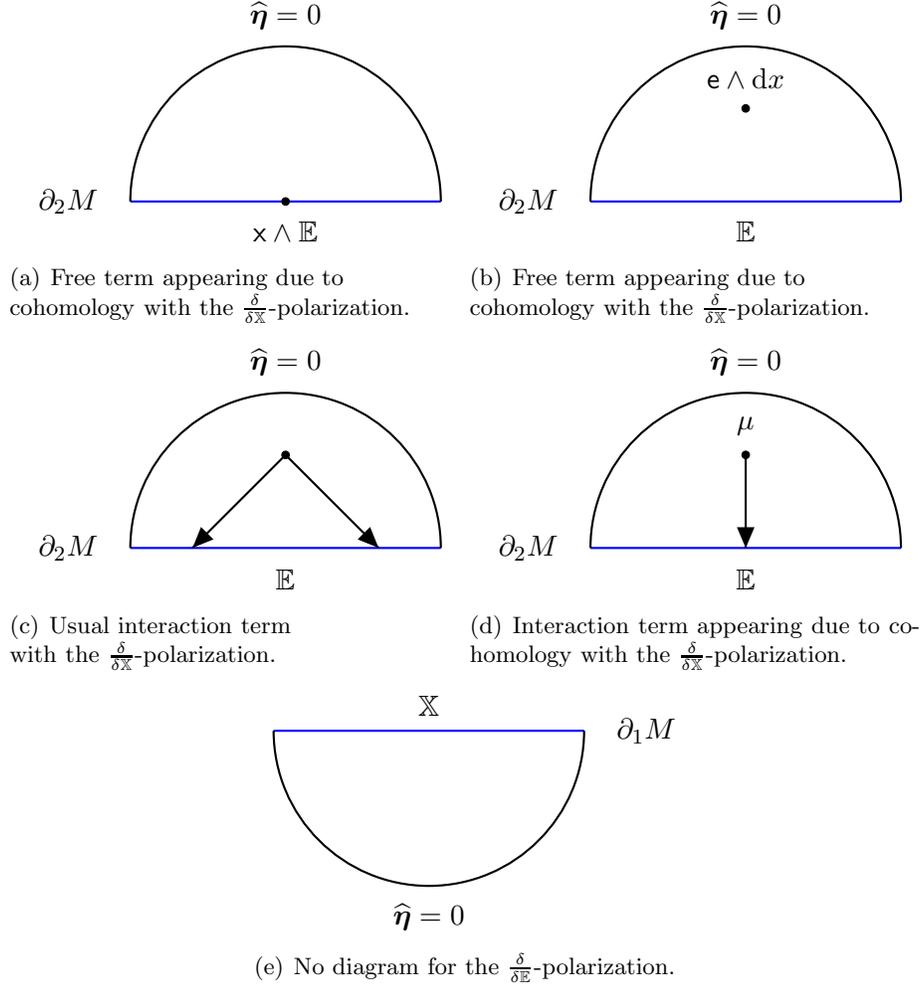

On $M=L_1$ one can choose the $\frac{\delta}{\delta\mathbb{X}}$- or a $\frac{\delta}{\delta\E}$-polarization, where the resulting Feynman diagrams are illustrated in figure \ref{L1}.

For figure \ref{subfigddEtrivial} we get the trivial state, denoted by $\widehat{\psi}_{L_1}^{\frac{\delta}{\delta\E}}$, since there is no diagram to compute. The perturbation term of the effective action for figure (a) is given by 
\[
\mathcal{S}^{\textnormal{pert,eff}}_{\partial_2 M}=\frac{1}{2}\alpha^{ij}\int_{C_2(\partial_2M)}\pi_{1}^*\E_i\land\widehat{\zeta}^\partial_{23}\land\pi_{2}^*\E_j,
\]
where $\widehat{\zeta}^\partial$ is again defined as in the setting of $L_3$. Now the free part contains also residual fields, since with the chosen polarization we get a non vanishing cohomology, namely $\mathcal{V}_{L_1} = H^{\bullet}(D) \oplus H^{\bullet}(D,\de D)$, generated by $1$ and $\mu$ respectively, where $\mu$ is a normalised volume form on $D$.  The two free terms in \ref{subfigfreexEddX} and \ref{subfigfreemuddX} sum to
$\mathcal{S}^{\textnormal{eff}}_{\partial_2M}=\int_{\partial_2M}z^{i}\mathbb{E}_{i}+z_i^\dagger \dd x^{i}$,
since the volume form on the bulk integrates to $1$. For the diagram \ref{subfigintmuddX} we get an additional perturbation term 
$
\widetilde{\mathcal{S}}^{\textnormal{pert,eff}}_{\partial_2M}=\alpha^{ij}\int_{\partial_2 M}z^{\dagger}_{i}\E_j\land\tau,
$
where $\{z^{i},z_i^\dagger\}$ are some chosen coordinates and dual coordinates on the cohomology and $\tau\in \Omega^1(\partial_2M)$ is the result of the integral over the bulk vertex of the graph with one bulk vertex connected to one boundary vertex with the property that 
\begin{equation}
\label{propagator1}
\dd\widehat{\zeta}^\partial=\pi_1^*\tau-\pi_2^*\tau.
\end{equation}
Therefore the overall state for $L_1$ is given by $\widehat{\psi}_{L_1}^{\frac{\delta}{\delta\mathbb{X}}}$. It is given by
\[
\widehat{\psi}_{L_1}=\widehat{\psi}_{L_1}^{\frac{\delta}{\delta\mathbb{X}}}=T_{L_1}\ee^{\frac{\I}{\hbar} \left(\mathcal{S}^{\textnormal{pert,eff}}_{\partial_2 M}+\mathcal{S}^{\textnormal{eff}}_{\partial_2M}+\tilde{\mathcal{S}}^{\textnormal{pert,eff}}_{\partial_2M}\right)}=T_{L_1}\ee^{\frac{\I}{\hbar}\mathcal{S}^{\textnormal{eff}}_{\partial M}}.
\]
Again, one can actually show that $T_{L_1}=1$, with respect to the basis of $\mathcal{V}_{L_1}$ given by $1$ and $\mu$. 

\section{The mdQME for the canonical relations}\label{s:mdQME}
\label{mdQMEsection}
Following the BV-BFV formalism, we need to make sure that the mdQME is satisfied for $\widehat{\psi}_{L_j}$, i.e.
\begin{equation}
\label{mdQME}
\nabla^{\partial L_j}_\textsf{G}\widehat{\psi}_{L_j}=\left(\dd +\I\hbar\triangle+\frac{\I}{\hbar}\Omega^{(j)}\right)\widehat{\psi}_{L_j}=0 
\end{equation}
for $j\in\{1,3\}$. Here the superscript $(j)$ means that $\Omega^{(j)}$ is the corresponding boundary BFV operator for $L_j$. This operator splits into a free part $\Omega_0^{(j)}$ and a perturbation part $\Omega_{\textnormal{pert}}^{(j)}$, i.e. $\Omega^{(j)}=\Omega_0^{(j)}+\Omega_{\textnormal{pert}}^{(j)}$. The mdQME will look different for different states, depending on the cohomology and the corresponding effective action. 

\subsection{The mdQME  for $L_3$}
By the general construction of \cite{CMR2}, one can check that $\Omega^{(3)}$ is given by the sum of  
\begin{align}
\label{bound_L3_1}
\Omega_{\textnormal{pert}}^{(3)}&=-\frac{1}{2}\alpha^{ij}\left(\hbar^2\int_{\partial_1M}\frac{\delta}{\delta\mathbb{X}_i}\frac{\delta}{\delta \mathbb{X}_j}-\int_{\partial_2^{(1)}M\sqcup\partial_2^{(2)}M}\E_i\land\E_j\right)\\
\label{bound_L3_2}
\Omega_0^{(3)}&=\I\hbar\left(\int_{\partial_1M} \dd\mathbb{X}_i\frac{\delta}{\delta\mathbb{X}_i}+\int_{\partial_2^{(1)}M\sqcup\partial_2^{(2)}M}\dd\E_i\frac{\delta}{\delta\E_i}+\int_{\partial_1M}\dd x^{i}\frac{\delta}{\delta\mathbb{X}_i}+\int_{\partial_2M}\E_i\land\dd x^{i}\right),
\end{align}

since $\widehat{\boldsymbol{\eta}}$ is quantized as $\widehat{\boldsymbol{\eta}}=-\I\hbar\frac{\delta}{\delta \textsf{X}}$. Also here one should note again the difference between the differentials of $\dd\mathbb{X}_i,\dd\mathbb{E}_i$ and $\dd x^{i}$. Since there is no cohomology for the chosen polarization, the mdQME reduces to 
$\Omega^{(3)}\widehat{\psi}_{L_3}=0$, which is indeed satisfied (see Appendix \ref{appendA} subsection \ref{comp_L3}).

\subsection{The mdQME  for $L_1$}
Equation \eqref{mdQME} needs to hold for $\widehat{\psi}_{L_1}$. Actually it only has to hold for $\widehat{\psi}^{\frac{\delta}{\delta\mathbb{X}}}_{L_1}$, which again represents the state for the diagrams with the $\frac{\delta}{\delta\mathbb{X}}$-polarization, since the mdQME for $\widehat{\psi}^{\frac{\delta}{\delta\E}}_{L_1}$ is trivially satisfied, we need only to check the case for $\widehat{\psi}_{L_1}^{\frac{\delta}{\delta\mathbb{X}}}$.
Indeed, if we denote by $\Omega^{(1)}_{\frac{\delta}{\delta\mathbb{X}}}$ the boundary BFV operator for $\widehat{\psi}^{\frac{\delta}{\delta\mathbb{X}}}_{L_1}$, then the mdQME for $\widehat{\psi}_{L_1}$ is equivalent with $\left(\hbar^2\triangle+\Omega^{(1)}_{\frac{\delta}{\delta\mathbb{X}}}\right)\widehat{\psi}_{L_1}^{\frac{\delta}{\delta\mathbb{X}}}=0$, where
\begin{align}
\Omega^{(1)}_{\frac{\delta}{\delta\mathbb{X}}}&=\int_{\partial_2M}\left(\I\hbar \dd\E_i\frac{\delta}{\delta\E_i}-\hbar^2\E_i \land\dd x^{i}+\frac{1}{2}\alpha^{ij}\E_i\land\E_j\right)\\
\triangle&=\sum_{i=1}^n(-1)^{1+\deg z^i}\frac{\partial}{\partial z^i}\frac{\partial}{\partial z^{\dagger}_i},
\end{align}
where $\deg z^{i}=1$ in our case. One can then check that the mdQME is indeed satisfied (see Appendix \ref{appendA} subsection \ref{comp_L1}) .

\section{Associativity and gluing}\label{s:assoc}
\label{associativity}

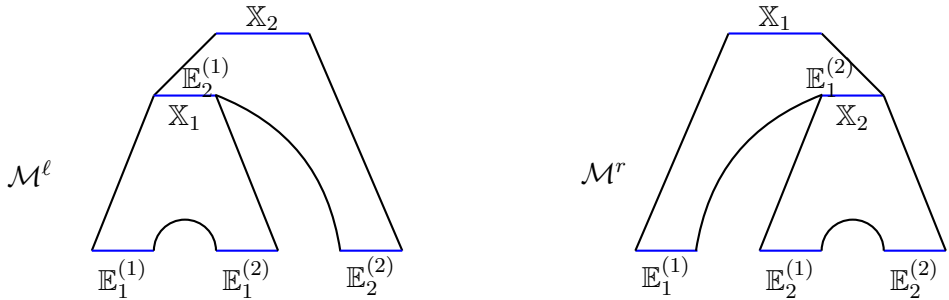
\begin{figure}[!ht]
\centering
\begin{minipage}{.5\textwidth}
\centering
\tikzset{
particle/.style={thick,draw=black},
particle2/.style={thick,draw=blue},
avector/.style={thick,draw=black, postaction={decorate},
    decoration={markings,mark=at position 1 with {\arrow[black]{triangle 45}}}},
gluon/.style={decorate, draw=black,
    decoration={coil,aspect=0}}
 }
\begin{tikzpicture}[x=0.05\textwidth, y=0.05\textwidth]
\draw[particle] (0,0)--(2,5) node[above]{};
\draw[particle] (2,5)--(4,7) node[right]{};
\draw[particle2] (2,5)--(4,5) node[below]{};
\draw[particle2] (4,7)--(7,7) node[above]{};
\draw[particle] (7,7)--(10,0) node[left]{};
\draw[particle2] (0,0)--(2,0) node[right]{};
\node[](u1) at (2,0){};
\node[](u2) at (4,0){};
\node[](u3) at (3.65,5.15){};
\node[](u4) at (8.055,-0.35){}; 
\draw[particle2] (4,0)--(6,0) node[right]{};
\draw[particle] (6,0)--(4,5) node[right]{};
\node[](1) at (3,4.3){$\mathbb{X}_1$};
\node[](11) at (3.7,5.6){$\mathbb{E}_2^{(1)}$};
\node[](2) at (1,0){};
\node[](2) at (5,0){};
\node[](e1) at (1,-0.9){$\E_1^{(1)}$};
\node[](e2) at (5,-0.9){$\E_1^{(2)}$};
\node[](x1) at (5.5,7.5){$\mathbb{X}_2$};
\node[](e4) at (9,-0.8){$\E_2^{(2)}$};
\node[](m) at (-2,2.5){$\mathcal{M}^\ell$};
\node[](e3) at (3,5.5){};
\node[](u5) at (10,0){};
\semiloop[particle]{u1}{u2}{0};
\draw[particle] (u4) to[bend right] (u3){}; 
\draw[particle2] (8,0)--(10,0){};
\end{tikzpicture}
\end{minipage}%
\begin{minipage}{.5\textwidth}
\centering
\tikzset{
particle/.style={thick,draw=black},
particle2/.style={thick,draw=blue},
avector/.style={thick,draw=black, postaction={decorate},
    decoration={markings,mark=at position 1 with {\arrow[black]{triangle 45}}}},
gluon/.style={decorate, draw=black,
    decoration={coil,aspect=0}}
 }
\begin{tikzpicture}[x=0.05\textwidth, y=0.05\textwidth]
\draw[particle] (0,0)--(2,5) node[above]{};
\draw[particle2] (-2,0)--(-4,0) node[right]{};
\draw[particle2] (2,5)--(4,5) node[right]{};
\draw[particle] (4,5)--(2,7) node[above]{};
\draw[particle2] (2,7)--(-1,7) node[above]{};
\draw[particle] (-1,7)--(-4,0) node[left]{};
\draw[particle2] (0,0)--(2,0) node[right]{};
\node[](u1) at (2,0){};
\node[](u2) at (4,0){};
\node[](u3) at (2.35,5.155){};
\node[](u4) at (-2.1,-0.33){}; 
\draw[particle2] (4,0)--(6,0) node[right]{};
\draw[particle] (6,0)--(4,5) node[right]{};
\node[](1) at (3,5){};
\node[](2) at (1,0){};
\node[](2) at (5,0){};
\node[](e1) at (1,-0.9){$\E_2^{(1)}$};
\node[](e2) at (5,-0.9){$\E_2^{(2)}$};
\node[](e3) at (2.3,5.6){$\E_1^{(2)}$};
\node[](e4) at (-3,-0.8){$\E_1^{(1)}$};
\node[](x1) at (3,4.3){$\mathbb{X}_2$};
\node[](x2) at (0.5,7.5){$\mathbb{X}_1$};
\node[](m) at (-5,2.5){$\mathcal{M}^r$};
\node[](u5) at (10,0){};
\semiloop[particle]{u1}{u2}{0};
\draw[particle] (u4) to[bend left] (u3){}; 
\end{tikzpicture}
\end{minipage}
\caption{The different ways of gluing two $L_3$s}
\label{assL_3}
\end{figure}

The next step includes the observation that the two different ways of gluing two $L_3$s as in figure \ref{assL_3} produce the same state up to some $\Omega^{(3)}$-exact term, which is responsible for the  associativity of the Moyal product out of the final gluing. To describe the state of the glued manifold as in the left figure, let $\Sigma^\ell$ be the identification of the boundary $\partial_2^{(1)}M$ for $M$ being the upper glued $L_3$, which we call $L_3^1$, and $\partial_1M$ for $M$ being the lower glued $L_3$, which we call $L_3^2$. 
 
Let us denote by $\mathcal{M}^{\ell}$ the left glued manifold, i.e. $\mathcal{M}^\ell:=L_3^{1}\cup_{\Sigma^\ell} L_3^2$. Following the gluing description of \cite{CMR2}, formally the state for $\mathcal{M}^\ell$ is given by the path integral
\begin{equation}
\label{glued_state1}
\widehat{\psi}_{\mathcal{M}^\ell}=\int_{\mathbb{X}_1^{\Sigma^\ell},\E^{\Sigma^\ell}_2} \ee^{-\frac{\I}{\hbar}\int_{\Sigma^\ell} \mathbb{X}^{\Sigma^\ell}_1\land \E^{\Sigma^\ell}_2}\widehat{\psi}_{L_3^1}\widehat{\psi}_{L_3^2}=\int_{\mathbb{X}_1^{\Sigma^\ell},\E^{\Sigma^\ell}_2} \ee^{-\frac{\I}{\hbar}\int_{\Sigma^\ell} \mathbb{X}^{\Sigma^\ell}_1\land \E^{\Sigma^\ell}_2}\ee^{\frac{\I}{\hbar}\left(\mathcal{S}^{\textnormal{eff}}_{\partial L_3^1}+\mathcal{S}^{\textnormal{eff}}_{\partial L_3^2}\right)}.
\end{equation}
The corresponding state for the right glued manifold $\mathcal{M}^r:=L_3^1\cup_{\Sigma^r} L_3^2$, where $\Sigma^r$ is the identification of the boundary component $\partial_2^{(2)}M$ for $M$ being the upper glued $L_3$, denoted by $L_3^1$ and $\partial_1M$ for $M$ being the lower $L_3$, denoted by $L_3^2$, is formally given by the path integral
\begin{equation}
\label{glued_state2}
\widehat{\psi}_{\mathcal{M}^r}=\int_{\mathbb{X}_2^{\Sigma^r},\E^{\Sigma^r}_1} \ee^{-\frac{\I}{\hbar}\int_{\Sigma^r} \mathbb{X}^{\Sigma^r}_2\land \E^{\Sigma^r}_1}\widehat{\psi}_{L_3^1}\widehat{\psi}_{L_3^2}=\int_{\mathbb{X}_2^{\Sigma^r},\E^{\Sigma^r}_1} \ee^{-\frac{\I}{\hbar}\int_{\Sigma^r} \mathbb{X}^{\Sigma^r}_1\land \E^{\Sigma^r}_2}\ee^{\frac{\I}{\hbar}\left(\mathcal{S}^{\textnormal{eff}}_{\partial L_3^1}+\mathcal{S}^{\textnormal{eff}}_{\partial L_3^2}\right)}.
\end{equation}

Instead of computing the glued states directly, one can consider a more general approach where associativity for our gluing actually appears as a special case. Therefore we need to describe the general manifold first.

\subsection{General associativity construction}

\begin{figure}[!ht]
\centering
\tikzset{
particle/.style={thick,draw=black},
particle2/.style={thick,draw=blue},
avector/.style={thick,draw=black, postaction={decorate},
    decoration={markings,mark=at position 1 with {\arrow[black]{triangle 45}}}},
gluon/.style={decorate, draw=black,
    decoration={coil,aspect=0}}
 }
\begin{tikzpicture}[x=0.03\textwidth, y=0.03\textwidth]
\node[](name) at (-3,2.5){$\mathcal{M}^n$};
\draw[particle] (-2,0)--(0,5) node[above]{};
\node[](int1) at (-1,-0.5){$I_1$};
\node[](int2) at (3,-0.5){$I_2$};
\node[](int3) at (13,-0.5){$I_{n-1}$};
\node[](int4) at (17,-0.5){$I_n$};
\node[](part1) at (8,5.5){$\partial_1\mathcal{M}^n$};
\draw[particle2] (-2,0)--(0,0) node[right]{};
\node[](u1) at (0,0){};
\node[](u2) at (2,0){};
\node[](u3) at (1.85,5.05){};
\node[](u4) at (6.05,-0.15){}; 
\node[](u5) at (4,0){}; 
\node[](u6) at (6,0){}; 
\node[](u7) at (10,0){};
\node[](u8) at (12,0){}; 
\node[](u9) at (14,0){}; 
\node[](u10) at (16,0){};
\draw[particle2] (2,0)--(4,0) node[right]{};
\node[](1) at (1,4.6){};
\node[](11) at (1.7,5.6){};
\node[](2) at (-1,0){};
\node[](2) at (3,0){};
\node[](e1) at (-1,-0.9){};
\node[](e2) at (3,-0.9){};
\node[](x1) at (3.5,7.4){};
\node[](e4) at (7,-0.8){};
\node[](m) at (-4,2.5){};
\node[](e3) at (1,5.5){};
\node[](u5) at (8,0){};
\semiloop[particle]{u1}{u2}{0};
\semiloop[particle]{u5}{u6}{0};
\draw[particle2] (6,0)--(7,0) node[below]{};
\node[](dots) at (8,0){$\hdots$};
\draw[particle2] (9,0)--(10,0) node[below]{};
\semiloop[particle]{u7}{u8}{0};
\draw[particle2] (12,0)--(14,0) node[below]{};
\semiloop[particle]{u9}{u10}{0};
\draw[particle2] (16,0)--(18,0) node[below]{};
\draw[particle] (18,0)--(16,5) node[below]{};
\draw[particle2] (16,5)--(0,5) node[below]{};
\end{tikzpicture}
\caption{The more general manifold $\mathcal{M}^n$.}
\label{gn}
\end{figure}
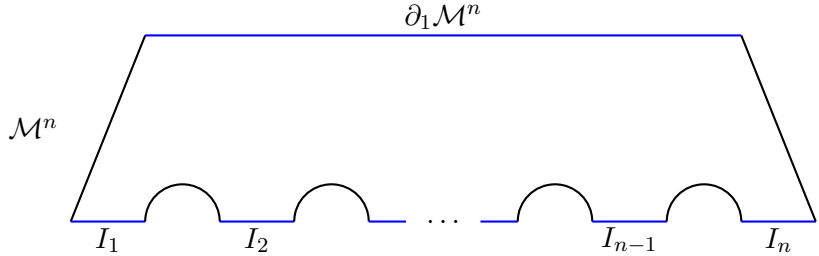

Let us consider the manifold $\mathcal{M}^n$ given as in figure \ref{gn}. Then we can describe
\[
\partial_2 \mathcal{M}^n=I_1\sqcup I_2\sqcup\dotsm \sqcup I_{n-1}\sqcup I_n
\]
as the disjoint union of the $n$ boundary components as in figure \ref{gn}. 

The state for $\mathcal{M}^n$ is now easily computed by considering the free and the interaction terms for the effective action similar as for $L_3$. Hence we get the effective action 
\[
\mathcal{S}_{\mathcal{M}^n}^{\textnormal{eff}}=-\int_{\partial_2\mathcal{M}^n\times \partial_1\mathcal{M}^n}\pi_1^*\E_i\land \zeta^n_{0\nu}\land\pi_2^*\mathbb{X}_i+\frac{1}{2}\alpha^{ij}\int_{\mathcal{M}^n\times C_2(\partial_2\mathcal{M}^n)}\zeta^n_{12}\land\zeta^n_{13}\land\pi_{2,1}^*\E_i\land\pi_{2,2}^*\E_j, 
\]
where $C_2(\partial_2\mathcal{M}^n)=C_2(I_1)\sqcup C_2(I_2)\sqcup\dotsm \sqcup I_1\times I_2\sqcup I_1\times I_3\sqcup\dotsm \sqcup I_{n-1}\times I_{n}$ and $\zeta^n$ is some bulk propagator for $\mathcal{M}^n$ (e.g. the one corresponding to the Euclidean metric). Now if  $\zeta^{n,'}$ is another bulk propagator on $\mathcal{M}^n$, since $\zeta^n$ is a propagator,  there is  (see \cite{CMR2} for the propagator construction or \cite{Mn} for the computation of deformation of a chain homotopy) a decomposition $\zeta^n=\zeta^{n,'}+\dd\kappa^n$, where  $\kappa^n$ is some zero form on $\mathcal{M}^n \times \mathcal{M}^n$. Generally, the difference of two propagators is given by an exact term for vanishing cohomology, whereas in the case of nonvanishing cohomology it is exact up to some terms depending on a basis of representatives for the cohomology classes. For our purpose, we consider a family $\zeta^{n,t}$ of these propagators by setting a parametrization given as $\zeta^{n,t}=\zeta+t\dd\kappa^n$ and look at the effective action for this family, which is given by
\begin{equation}
\mathcal{S}_{\mathcal{M}^n}^{\textnormal{eff}}(t)=-\int_{\partial_2\mathcal{M}^n\times \partial_1\mathcal{M}^n}\pi_1^*\E_i\land \zeta^{n,t}_{0\nu}\land\pi_2^*\mathbb{X}_i+\frac{1}{2}\alpha^{ij}\int_{\mathcal{M}^n\times C_2(\partial_2\mathcal{M}^n)}\zeta^{n,t}_{12}\land\zeta^{n,t}_{13}\land\pi_{2,1}^*\E_i\land\pi_{2,2}^*\E_j.
\end{equation}
The state for this parametrization is thus given by
\begin{equation}
\label{gen_state}
\widehat{\psi}_{\mathcal{M}^n}(t)=T_{\mathcal{M}^n}\ee^{\frac{\I}{\hbar}\mathcal{S}_{\mathcal{M}^n}^{\textnormal{eff}}(t)}.
\end{equation}
Let us set 
\begin{multline*}
\mathcal{A}=\int_{\partial_2\mathcal{M}^n\times \partial_1\mathcal{M}^n}\pi^*_1\E_i\land \kappa^n_{0\nu}\land \pi_2^*\mathbb{X}_i+\frac{1}{2}\alpha^{ij}\int_{\mathcal{M}^n\times C_2(\partial_2\mathcal{M}^n)}\zeta_{12}^{n,t}\land\kappa^n_{13}\land\pi_{2,1}^*\E_i\land\pi_{2,2}^*\E_j\\+\frac{1}{2}\alpha^{ij}\int_{\mathcal{M}^n\times C_2(\partial_2\mathcal{M}^n)}\kappa_{12}^n\land\zeta_{13}^{n,t}\land\pi_{2,1}^*\E_i\land\pi_{2,2}^*\E_j.
\end{multline*}
Then we get that (see Appendix \ref{comp_assoc})
\begin{equation}
\label{time_vary}
\partial_t\widehat{\psi}_{\mathcal{M}^n}(t)=\Omega^{(3)}\left(\widehat{\psi}_{\mathcal{M}^n}(t)\mathcal{A}\right),
\end{equation}
By taking $n = 3$ and $\zeta^n,\zeta^{n,'}$ the propagators\footnote{Recall from \cite{CMR2} that the gluing of two propagators along a common boundary is again a propagator for the glued manifold.} arising from the two different gluings of $\mathcal{M}^3$ this provides a general way of showing that associativity is indeed satisfied, namely, since $\widehat{\psi}_{\mathcal{M}^n}$ changes by an $\Omega^{(3)}$-exact term, we can say that $\widehat{\psi}_{\mathcal{M}^\ell}-\widehat{\psi}_{\mathcal{M}^3}$ is given by some $\Omega^{(3)}$-exact term, say $\Omega^{(3)}(B)$ and $\widehat{\psi}_{\mathcal{M}^r}-\widehat{\psi}_{\mathcal{M}^3}$ is also given by some other $\Omega^{(3)}$-exact term, say $\Omega^{(3)}(A)$. Hence we can say that $\widehat{\psi}_{\mathcal{M}^\ell}$ and $\widehat{\psi}_{\mathcal{M}^r}$ also differ by an $\Omega^{(3)}$-exact term since 
$\widehat{\psi}_{\mathcal{M}^\ell}-\widehat{\psi}_{\mathcal{M}^3}+\widehat{\psi}_{\mathcal{M}^3}-\widehat{\psi}_{\mathcal{M}^r}=\Omega^{(3)}(A+B)$.

\section{Main gluing and the Moyal product}\label{s:main}
 
\subsection{Flat observables on disks}
To retrieve the star product we need to include boundary observables, namely those that are induced by functions on $\mathscr{P}$. To such a function and point $u_0$ which lies in some interval with $\widehat{\boldsymbol\eta}= 0$ boundary condition we can associate the observable $X \mapsto f(X(u_0))$. The expectation value of such an observable is a section of $\calH^{tot}$ which is constant on boundary fields. Choose coordinates $\{z,z^\dagger\}$ on the space of residual fields (if it is nonempty, cf the discussion in \ref{Residual fields}). The result in section \ref{4.6} gives us a Grothendieck connection $D_{\textsf{G}}$ on $\Gamma(\widehat{S}T^*\mathscr{P})$ as
\[
D_{\textsf{G}}=\sum_{j=1}^d\dd x^{j}\left(\frac{\partial}{\partial x^{j}}-\frac{\partial}{\partial z^j}\right),
\]
Hence for a smooth function $f\in C^\infty(\mathscr{P})$ we get that 
\[
D_{\textsf{G}}f(x+z)=0,
\]
i.e. functions on $T^*\mathscr{P}$ of this form are flat sections with respect to the connection $D_{\textsf{G}}$.

%

\subsection{First example of gluing}
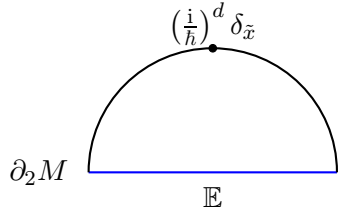
\begin{figure}[!ht] 
\centering
 \tikzset{
particle/.style={thick,draw=black},
particle2/.style={thick,draw=blue},
avector/.style={thick,draw=black, postaction={decorate},
    decoration={markings,mark=at position 1 with {\arrow[black]{triangle 45}}}},
gluon/.style={decorate, draw=black,
    decoration={coil,aspect=0}}
 }
\begin{tikzpicture}[x=0.04\textwidth, y=0.04\textwidth]
\draw[particle2] (0,-2.5)--(5,-2.5) node[above]{};
\node[](u1) at (0,-2.5){};
\node[](u2) at (5,-2.5){};
\node[](2) at (2.5,0){};
\draw[fill=black] (2) circle (0.05cm);
\node[](eta1) at (2.5,-3){$\E$};
\node[](eta2) at (2.5,0.5){$\left(\frac{\I}{\hbar}\right)^d\delta_{\tilde x}$};
\semiloop[particle]{u1}{u2}{0};
\node(n) at (-1,-2.5){$\partial_2M$};
\end{tikzpicture}
\caption{$L_1$ with the $\frac{\delta}{\delta\mathbb{X}}$-polarization and endowed with a delta function}
\label{delta1}
\end{figure}

We will now present the actual gluing and modification that is needed to recover the Moyal product. Let us therefore first take a look at $L_1$ with the $\frac{\delta}{\delta\mathbb{X}}$-polarization and the difference that we attach a delta function $\left(\frac{\I}{\hbar}\right)^d\delta_{\tilde x}$, defined on $\mathscr{P}$, on the black boundary component where we have set $\widehat{\boldsymbol{\eta}}=0$ before, as it is shown in figure \ref{delta1}. The renormalization constant in front of the delta function will become clear in the following.
 
Here $\delta_{\tilde x}(x)=\delta(x-\tilde x)$, where $\tilde x$ and $x$ are points in $\mathscr{P}$. 
Hence we can obtain the state for this modified $L_1$ manifold, denoted by $L_1^{\left(\frac{\I}{\hbar}\right)^d\delta_{\tilde x}}$, as 
\[
\widehat{\psi}_{L^{\left(\frac{\I}{\hbar}\right)^d\delta_{\tilde x}}_1}(\E,z,z^\dagger,x,\dd x)=\left(\frac{\I}{\hbar}\right)^d\delta_{\tilde x}(x+z)\ee^{\frac{\I}{\hbar}\left(z^{i}\int_{\partial_2M}\E_i+z^\dagger_j\dd x^{j}\right)}.
\]
We now want to consider a special gluing, which we split into two cases of different structure.

\subsubsection{First case}

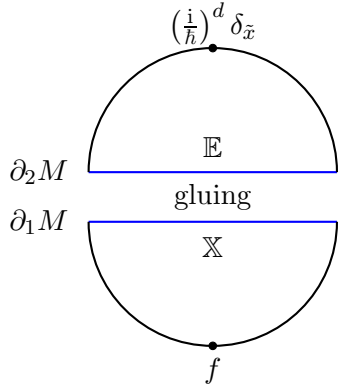
\begin{figure}[!ht] 
\centering
 \tikzset{
particle/.style={thick,draw=black},
particle2/.style={thick,draw=blue},
avector/.style={thick,draw=black, postaction={decorate},
    decoration={markings,mark=at position 1 with {\arrow[black]{triangle 45}}}},
gluon/.style={decorate, draw=black,
    decoration={coil,aspect=0}}
 }
\begin{tikzpicture}[x=0.04\textwidth, y=0.04\textwidth]
\draw[particle2] (0,-2.5)--(5,-2.5) node[above]{};
\draw[particle2] (0,-3.5)--(5,-3.5) node[above]{};
\node[](u1) at (0,-2.5){};
\node[](u2) at (5,-2.5){};
\node[](u3) at (0,-3.5){};
\node[](u4) at (5,-3.5){};
\node[](2) at (2.5,0){};
\node[](3) at (-1,-3.5){$\partial_1M$};
\node[](4) at (2.5,-4){$\mathbb{X}$};
\node[](5) at (2.5,-6.5){$f$};
\node[](6) at (2.5,-6){};
\node[](g) at (2.5,-3){gluing};
\draw[fill=black] (2) circle (0.05cm);
\draw[fill=black] (6) circle (0.05cm);
\node[](eta1) at (2.5,-2){$\E$};
\node[](eta2) at (2.5,0.5){$\left(\frac{\I}{\hbar}\right)^d\delta_{\tilde x}$};
\semiloop[particle]{u4}{u3}{180};
\semiloop[particle]{u1}{u2}{0};
\node(n) at (-1,-2.5){$\partial_2M$};
\end{tikzpicture}
\caption{The gluing of $L_1^{\left(\frac{\I}{\hbar}\right)^d\delta_{\tilde x}}$ and $L_1^f$}
\label{gluing_L1}
\end{figure}

Let us consider a first important gluing for $\alpha=0$, that is that of $L_1^{\left(\frac{\I}{\hbar}\right)^d\delta_{\tilde x}}$ with an $L_1$ manifold endowed with the $\frac{\delta}{\delta\mathbb{E}}$-polarization, where we attach a smooth function $f\in C^\infty(\mathscr{P})$ at the black boundary component, where we have set $\widehat{\boldsymbol{\eta}}=0$ before, which we denote by $L_1^f$. Moreover, we glue along the boundary components where the fields are attached as it is shown in figure \ref{gluing_L1}. 

We expect to end up with the observable $f$ using this particular gluing. Indeed, first we notice that the Feynman diagrams that we get for $L_1^{f}$ are all possible arrows going from $\partial_1M$ to the observable $f$. Hence, by Wick's theorem, we get the state
\[
\widehat{\psi}_{L_1^f}=\sum_{n\geq 0}\frac{(-\I\hbar)^n}{n!}\int_{C_n(\partial_1M)}\mathbb{X}^{i_1}(u_1)\dotsm\mathbb{X}^{i_n}(u_n)\zeta(u_1,v_0)\dotsm\zeta(u_n,v_0)\partial_{i_1}\dotsm\partial_{i_n}f(x),
 \]
where $u_1,...,u_n$ are points on $\partial_1M$ and $v_0$ a point on the lower boundary component, where $f$ is attached and $\mathbb{X}^{i_j}$ are the component fields of $\mathbb{X}$ on $\partial_1M$. Using the fact that $\int_{\partial_1M}\zeta(u_j,v_0)=1$ for all $j=1,...,n$, we can use the gluing principle of identifying the $\mathbb{X}$- with the $\E$-fields to obtain the glued state
\[
\widehat{\psi}_{f,\tilde x}=\left(\frac{\I}{\hbar}\right)^d\delta_{\tilde x}(x+z)\sum_{n\geq 0}\frac{1}{n!}z^{i_1}\dotsm z^{i_n}\partial_{i_1}\dotsm\partial_{i_n}f(x)\ee^{\frac{\I}{\hbar} z^\dagger_j\dd x^{j}},
 \]
 which can be written as
 \[
 \widehat{\psi}_{f,\tilde x}=\left(\frac{\I}{\hbar}\right)^d\delta_{\tilde x}(x+z)f(x+z)\ee^{\frac{\I}{\hbar} z^\dagger_j\dd x^{j}},
 \]
by considering the sum as a Taylor expansion of $f$. Let us now write 
\[
\widehat{\psi}_{f,\tilde x}=\widehat{\psi}_{\rho}=\rho(x+z)\ee^{\frac{\I}{\hbar} z^\dagger_j\dd x^{j}},
\]
with $\rho(x):=\left(\frac{\I}{\hbar}\right)^d\delta(x-\tilde x)f(x)$. Thus we get 
\begin{align*}
\triangle\widehat{\psi}_{\rho}&=\frac{\I}{\hbar}\sum_{k=1}^d\dd x^k\partial_k\rho(x+z) \ee^{\frac{\I}{\hbar} z^\dagger_j\dd x^{j}},\\
 \dd\widehat{\psi}_{\rho}&=\sum_{k=1}^d \dd x^k\partial_k\rho(x+z)\ee^{\frac{\I}{\hbar} z^\dagger_j\dd x^{j}},
 \end{align*}
 and therefore $\triangle\widehat{\psi}_{\rho}=\frac{\I}{\hbar}\dd\widehat{\psi}_{\rho}$. Moreover, since $\Omega\widehat{\psi}_\rho=0$ we get that the mdQME holds. Let us now consider the Lagrangian submanifold $\mathcal{L}=\{z=0\}$ and the corresponding BV integral 
 \[
 \int_{\mathcal{L}}\widehat{\psi}_{\rho}\Big|_{z=0}\dd z^\dagger_1\dotsm \dd z^\dagger_d=\rho(x)\left(\frac{\hbar}{\I}\right)^d\dd^dx=f(x)\delta(x-\tilde x)\dd^dx.
 \]
Now integrating this term over the whole Poisson manifold, we get 
\[
\int_\mathscr{P}\left(\int_{\mathcal{L}}\widehat{\psi}_\rho\Big|_{z=0}\dd z^\dagger_1\dotsm \dd z^\dagger_d\right)=f(\tilde x),
 \]
and therefore we end up with our function again.

\subsubsection{Second case}
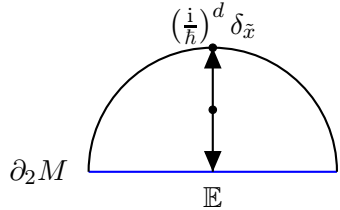
\begin{figure}[!ht]
\centering
\tikzset{
particle/.style={thick,draw=black},
particle2/.style={thick,draw=blue},
avector/.style={thick,draw=black, postaction={decorate},
    decoration={markings,mark=at position 1 with {\arrow[black]{triangle 45}}}},
gluon/.style={decorate, draw=black,
    decoration={coil,aspect=0}}
 }
\begin{tikzpicture}[x=0.04\textwidth, y=0.04\textwidth]
\draw[particle2] (0,-2.5)--(5,-2.5) node[above]{};
\node[](u1) at (0,-2.5){};
\node[](u2) at (5,-2.5){};
\node[](2) at (2.5,-1.25){};
\node[](3) at (2.5,0){};
\draw[fill=black] (2) circle (0.05cm);
\draw[fill=black] (3) circle (0.05cm);
\draw[avector] (2.5,-1.25)--(2.5,-2.5) node[above]{};
\draw[avector] (2.5,-1.25)--(2.5,0) node[above]{};
\node[](eta1) at (2.5,-3){$\E$};
\node[](eta2) at (2.5,0.5){$\left(\frac{\I}{\hbar}\right)^d\delta_{\tilde x}$};
\semiloop[particle]{u1}{u2}{0};
\node(n) at (-1,-2.5){$\partial_2M$};
\end{tikzpicture}
\caption{The additional diagram for $L_1^{\left(\frac{\I}{\hbar}\right)^d\delta_{\tilde x}}$}
\label{delta}
\end{figure}

If we now consider $\alpha$ to be constant different from zero, we can do the same computation with the difference that we have an additional perturbation term $\alpha^{ij}\int_{\partial_1M}z^{i}\E_i\land \tau$ in the exponential for the state $\widehat{\psi}_{L^{\left(\frac{\I}{\hbar}\right)^d\delta_{\tilde x}}_1}$. In fact we have now the corresponding diagrams as in figure \ref{L1} (a) and (c) with the additional diagram for the $\frac{\delta}{\delta\mathbb{X}}$-polarization, where we have one bulk vertex with one arrow going to the boundary $\partial_2M$ and with one arrow deriving the function $\left(\frac{\I}{\hbar}\right)^d\delta_{\tilde x}$ as in figure \ref{delta}.

Therefore one obtains that the glued state $\widehat{\psi}_{f,\tilde x}$ is actually given by the star product $f\star_M\left(\frac{\I}{\hbar}\right)^d\delta_{\tilde x}$. To consider all points on $\mathscr{P}$, we integrate over all background fields to obtain
\begin{align*}
\int_{\mathscr{P}}\left(f\star_M \left(\frac{\I}{\hbar}\right)^d\delta_{\tilde x}\right)(x) \dd^dx&=\underbrace{\int_\mathscr{P}\left(\frac{\I}{\hbar}\right)^df(x)\delta(x-\tilde x)\dd^dx}_{=f(\tilde x)}\\
&+\underbrace{\left(\frac{\I}{\hbar}\right)^d\alpha^{ij}\int_{\mathscr{P}}\partial_if(x)\partial_j\delta(x-\tilde x)\dd^dx}_{=\underbrace{\left(\frac{\I}{\hbar}\right)^d\alpha^{ij}\int_\mathscr{P}\partial_i\partial_j f(x)\delta(x-\tilde x)\dd^dx}_{=0}+\dotsm}+\dotsm
 \end{align*}
where we have used the fact that $\alpha$ is an antisymmetric tensor to obtain zero for the other terms. Therefore we end up with the function $f$ for the gluing of $L_1^{\delta_{\tilde x}}$ and $L_1^f$. For notational symplicity we have only delt with the case where $f$ is analytical, but the final result is general.

\begin{rem}
One should note that the computations at the end can also be done with general Poisson structures, in particular with Kontsevich's star product.
\end{rem}

\subsection{Obtaining the Moyal product} 
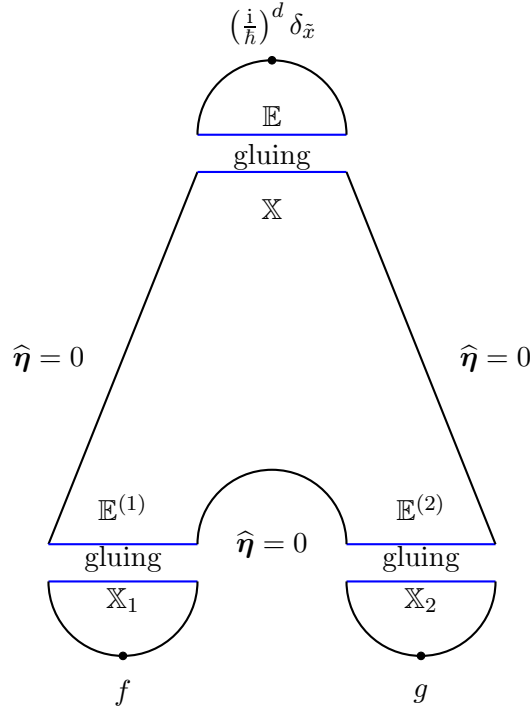
\begin{figure}[!ht]
\centering
\tikzset{
particle/.style={thick,draw=black},
particle2/.style={thick,draw=blue},
avector/.style={thick,draw=black, postaction={decorate},
    decoration={markings,mark=at position 1 with {\arrow[black]{triangle 45}}}},
gluon/.style={decorate, draw=black,
    decoration={coil,aspect=0}}
 }
\begin{tikzpicture}[x=0.06\textwidth, y=0.06\textwidth]
\draw[particle] (0,0)--(2,5) node[above]{};
\draw[particle2] (2,5.5)--(4,5.5)node[above]{};
\draw[particle2] (2,5)--(4,5) node[right]{};
\draw[particle2] (0,0)--(2,0) node[right]{};
\draw[particle2] (0,-0.5)--(2,-0.5)node[below]{};
\draw[particle2] (4,-0.5)--(6,-0.5)node[below]{};
\node[](u1) at (2,0){};
\node[](u2) at (4,0){};
\draw[particle2] (4,0)--(6,0) node[right]{};
\draw[particle] (6,0)--(4,5) node[right]{};
\node[](2) at (0.7,0){};
\node[](3) at (1.4,0){};
\node[](11) at (2,5.5){};
\node[](12) at (4,5.5){};
\node[](13) at (0,-0.5){};
\node[](14) at (2,-0.5){};
\node[](15) at (4,-0.5){};
\node[](16) at (6,-0.5){};
\node[](t1) at (3,5.2){gluing};
\node[](t2) at (1,-0.2){gluing};
\node[](t2) at (5,-0.2){gluing};
\node[](p1) at (3,6.5){};
\node[](p2) at (1,-1.5){};
\node[](p3) at (5,-1.5){};
\node[](p4) at (3,7){$\left(\frac{\I}{\hbar}\right)^d\delta_{\tilde x}$};
\draw[fill=black] (p1) circle (0.05cm);
\draw[fill=black] (p2) circle (0.05cm);
\draw[fill=black] (p3) circle (0.05cm);
\node[](f) at (1,-2){$f$};
\node[](g) at (5,-2){$g$};
\node[](x13) at (5,-0.75){$\mathbb{X}_2$};
\node[](x12) at (1,-0.75){$\mathbb{X}_1$};
\node[](eta11) at (3,5.75){$\E$};
\node[](eta1) at (0,2.5){$\widehat{\boldsymbol{\eta}}=0$};
\node[](eta2) at (6,2.5){$\widehat{\boldsymbol{\eta}}=0$};
\node[](eta3) at (3,0){$\widehat{\boldsymbol{\eta}}=0$};
\node[](e1) at (1,0.5){$\E^{(1)}$};
\node[](e2) at (5,0.5){$\E^{(2)}$};
\node[](e3) at (3,4.5){$\mathbb{X}$};
\semiloop[particle]{u1}{u2}{0};
\semiloop[particle]{11}{12}{0};
\semiloop[particle]{14}{13}{180};
\semiloop[particle]{16}{15}{180};
\end{tikzpicture}
\caption{The gluing for the Moyal product}
\label{gluing_moyal}
\end{figure}

Now using this construction, we can recover the Moyal product by the gluing shown in figure \ref{gluing_moyal}. 
 
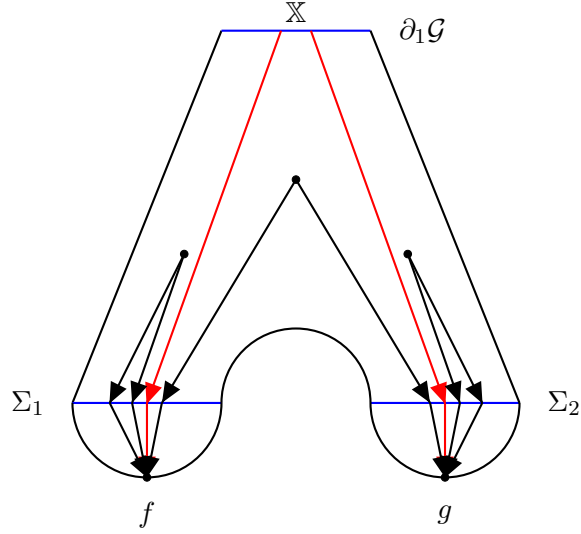
\begin{figure}[!ht]
\centering
\tikzset{
particle/.style={thick,draw=black},
particle2/.style={thick,draw=blue},
avector/.style={thick,draw=black, postaction={decorate},
    decoration={markings,mark=at position 1 with {\arrow[black]{triangle 45}}}},
avector2/.style={thick,draw=red, postaction={decorate},
    decoration={markings,mark=at position 1 with {\arrow[red]{triangle 45}}}},
gluon/.style={decorate, draw=black,
    decoration={coil,aspect=0}}
 }
\begin{tikzpicture}[x=0.06\textwidth, y=0.06\textwidth]
\draw[particle] (0,0)--(2,5) node[above]{};
\draw[particle2] (2,5)--(4,5) node[right]{};
\draw[particle2] (0,0)--(2,0) node[right]{};
\node[](u1) at (2,0){};
\node[](u2) at (4,0){};
\node[](X) at (3,5.25){$\mathbb{X}$};
\node[](boundary) at (4.7,5){$\partial_1\mathcal{G}$};
\draw[particle2] (4,0)--(6,0) node[right]{};
\draw[particle] (6,0)--(4,5) node[right]{};
\node[](2) at (0.7,0){};
\node[](3) at (1.4,0){};
\node[](11) at (2,5.5){};
\node[](12) at (4,5.5){};
\node[](13) at (0,0){};
\node[](14) at (2,0){};
\node[](15) at (4,0){};
\node[](16) at (6,0){};
\node[](p2) at (1,-1){};
\node[](p3) at (5,-1){};
\node[](n1) at (1.5,2){};
\node[](n2) at (3,3){};
\node[](n3) at (4.5,2){};
\node[](n4) at (2.5,5){};
\node[](n5) at (3.5,5){};
\draw[avector2] (2.8,5)--(1,0);
\draw[avector2] (1,0)--(1,-1);
\draw[avector2] (3.2,5)--(5,0);
\draw[avector2] (5,0)--(5,-1);
\draw[avector] (1.5,2)--(0.5,0);
\draw[avector] (1.5,2)--(0.8,0);
\draw[avector] (0.5,0)--(1,-1);
\draw[avector] (0.8,0)--(1,-1);
\draw[avector] (3,3)--(1.2,0);
\draw[avector] (1.2,0)--(1,-1);
\draw[avector] (3,3)--(4.8,0);
\draw[avector] (4.8,0)--(5,-1);
\draw[avector] (4.5,2)--(5.5,0);
\draw[avector] (5.5,0)--(5,-1);
\draw[avector] (4.5,2)--(5.2,0);
\draw[avector] (5.2,0)--(5,-1);
\draw[fill=black] (n1) circle (0.05cm);
\draw[fill=black] (n2) circle (0.05cm);
\draw[fill=black] (n3) circle (0.05cm);
\draw[fill=black] (p2) circle (0.05cm);
\draw[fill=black] (p3) circle (0.05cm);
\node[](f) at (1,-1.5){$f$};
\node[](sigma1) at (-0.6,0){$\Sigma_1$};
\node[](sigma2) at (6.6,0){$\Sigma_2$};
\node[](g) at (5,-1.5){$g$};
\semiloop[particle]{u1}{u2}{0};
\semiloop[particle]{14}{13}{180};
\semiloop[particle]{16}{15}{180};
\end{tikzpicture}
\caption{Appearing diagrams, where $\Sigma_1$ and $\Sigma_2$ are corresponding identified boundaries for the gluing. The red arrows represent the free part terms. We can also observe that by antisymmetry the two diagrams, where two interaction term arrows go to the same glued boundary component, vanish. Thus the Moyal product of $f$ and $g$ will appear in the glued result for the interaction diagram going in both directions.}
\label{moyal_diag}
\end{figure}

\begin{figure}[!ht]
\centering
\tikzset{
particle/.style={thick,draw=black},
particle2/.style={thick,draw=blue},
avector/.style={thick,draw=black, postaction={decorate},
    decoration={markings,mark=at position 1 with {\arrow[black]{triangle 45}}}},
gluon/.style={decorate, draw=black,
    decoration={coil,aspect=0}}
 }
\begin{tikzpicture}[x=0.04\textwidth, y=0.04\textwidth]
\draw[particle2] (0,0)--(5,0) node[above]{};
\node[](u1) at (0,0){};
\node[](u2) at (5,0){};
\node[](2) at (2.5,-2){};
\node[](eta1) at (2.5,0.4){$\mathbb{X}$};
\node[](eta2) at (2.5,-3){$f\star_M g$};
\draw[fill=black] (2.5,-2.5) circle (0.05cm);
\semiloop[particle]{u2}{u1}{180};
\end{tikzpicture}
\caption{The manifold $L_1^{f\star_Mg}$}
\label{mL_1}
\end{figure}
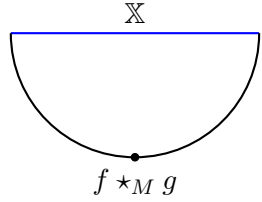

According to the notation $L_1^f$, we consider the same manifold but with $f\star_M g$ attached to it when we write $L_1^{f\star_Mg}$ (see figure \ref{mL_1}) for any $f,g\in C^\infty(\mathscr{P})$. The idea of the gluing in figure \ref{gluing_moyal} is to produce the same result as if one would do the gluing of $L_1^{\left(\frac{\I}{\hbar}\right)^d\delta_{\tilde x}}$ and $L_1^{f\star_Mg}$ as before. Therefore we first compute the state of the appearing Feynman diagrams for the partial gluing as in figure \ref{moyal_diag}, by gluing $L_1^f$ and $L_1^g$ to $\partial_2^{(1)}L_3$ and $\partial_2^{(2)}L_3$. The diagrams in figure \ref{moyal_diag} are those which need to be evaluated under the exponential map to obtain the state of the glued manifold in figure \ref{moyal_diag}, which we denote by $\mathcal{G}$.

Using Wick's theorem, the effective action terms of $L_3$ and the gluing process as in \cite{CMR2}, we can compute the state of the glued manifold $\mathcal{G}$ by the same arguments as before. Thus we get

\begin{multline}
\label{finalstate}
\widehat{\psi}_{\mathcal{G}}=\sum_{n,m,\ell\geq 0}\frac{(-\I\hbar)^{n+m+\ell}}{n!m!\ell !}\int_{C_{n+m}(\partial_1\mathcal{G})} \prod_{k_1=1}^n\prod_{k_2=1}^m\mathbb{X}^{i_{k_1}}(u_{k_1})\zeta(u_{k_1},v_0)\mathbb{X}^{j_{k_2}}(\tilde{u}_{k_2})\zeta(\tilde{u}_{k_2},v_1)\\\times\prod_{k_3=1}^\ell\alpha^{i_{k_3}j_{k_3}}\partial_{i_{k_1}}\partial_{i_{k_3}}f(x)\partial_{j_{k_2}}\partial_{j_{k_3}}g(x),
\end{multline}
where $\partial_1\mathcal{G}$ is the top boundary, $u_1,...,u_n,\tilde{u}_1,...,\tilde{u}_m$ are distinct points in $\partial_1\mathcal{G}$, the $\mathbb{X}^{i_k}$-fields are the component fields of the $\mathbb{X}$-field on $\partial_1\mathcal{G}$ corresponding to the $\Sigma_1$ gluing and the $\mathbb{X}^{j_k}$-fields are those corresponding to the $\Sigma_2$ gluing. Here the propagators are given as before with the difference that $v_0$ represents a point on the lower boundary component of $L_1^f$ where $f$ is attached and $v_1$ represents a point on the lower boundary component of $L_1^g$ where $g$ is attached. 

\begin{figure}[!ht] 
\centering
 \tikzset{
particle/.style={thick,draw=black},
particle2/.style={thick,draw=blue},
avector/.style={thick,draw=black, postaction={decorate},
    decoration={markings,mark=at position 1 with {\arrow[black]{triangle 45}}}},
gluon/.style={decorate, draw=black,
    decoration={coil,aspect=0}}
 }
\begin{tikzpicture}[x=0.04\textwidth, y=0.04\textwidth]
\draw[particle2] (0,-2.5)--(5,-2.5) node[above]{};
\draw[particle2] (0,-3.5)--(5,-3.5) node[above]{};
\node[](u1) at (0,-2.5){};
\node[](u2) at (5,-2.5){};
\node[](u3) at (0,-3.5){};
\node[](u4) at (5,-3.5){};
\node[](2) at (2.5,0){};
\node[](3) at (-1,-3.5){$\partial_1M$};
\node[](4) at (2.5,-4){$\mathbb{X}$};
\node[](5) at (2.5,-6.5){$f\star_M g$};
\node[](6) at (2.5,-6){};
\node[](g) at (2.5,-3){gluing};
\draw[fill=black] (2) circle (0.05cm);
\draw[fill=black] (6) circle (0.05cm);
\node[](eta1) at (2.5,-2){$\E$};
\node[](eta2) at (2.5,0.5){$\left(\frac{\I}{\hbar}\right)^d\delta_{\tilde x}$};
\semiloop[particle]{u4}{u3}{180};
\semiloop[particle]{u1}{u2}{0};
\node(n) at (-1,-2.5){$\partial_2M$};
\end{tikzpicture}
\caption{The cap gluing for the Moyal product}
\label{gluing2_L1}
\end{figure}
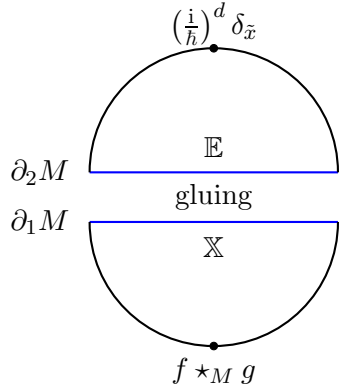

Now by the argument as before, we can consider the gluing as in figure \ref{gluing2_L1}, by identifying the states $\widehat{\psi}_{L_1^{f\star_Mg}}$ and $\widehat{\psi}_\mathcal{G}$ with each other by applying the product rule, only with the difference that we have two different evaluation points for $\widehat{\psi}_\mathcal{G}$. Then we can glue these states together as in figure \ref{gluing2_L1}, which also corresponds then to the gluing of figure \ref{gluing_moyal}.

Finally, using the procedure as before for such a gluing, we end up with the integral 
\[
\int_{\mathscr{P}}\left\{(f\star_M g)\star_M\left(\frac{\I}{\hbar}\right)^d\delta_{\tilde x}\right\}(x)\dd^dx,
\]
which yields $f\star_M g(\tilde x)$. Therefore we obtain the Moyal product of $f$ and $g$ as claimed. Moreover, we have already discussed that associativity is satisfied, since the states of the $L_3$ gluings only differ by an $\Omega^{(3)}$-exact term, i.e. the mdQME is directly satisfied and therefore associativity holds for this construction.

\section{Outlook}
 
The construction of the present paper may be generalized to other Poisson structures.

We now briefly list the main peculiarities of  the general case.
\begin{itemize}
\item If the target Poisson manifold is not an open subset of $\R^d$, the Grothendieck connection does not
have the simple form used in equation \eqref{action}. A more general form is available, see \cite{BCM}, and this will lead
to changes in the boundary action.
\item The boundary operator $\Omega$ is not as simple as in the case of a constant Poisson structure. Still 
following \cite{CMR2} it can be computed explicitly, it squares to zero, the states solve the mQME and, under a change of gauge fixing,
change by an exact term for the connection $\nabla^{\partial\Sigma}_{\textsf{G}}$.
\item The computation of the states $L_1$ and $L_3$ requires many more (in general infinitely many) Feynman diagrams. As a consequence the state corresponding to a function $f$ on the Poisson manifold has 
perturbative corrections.
Still, by the general results of \cite{CMR2}, these states satisfiy the mQME and change accordingly by a change of gauge fixing. In particular, $L_3$ still defines a product that is associative up to exact terms for the connection
$\nabla^{\partial\Sigma}_{\textsf{G}}$.
\item By the general results of \cite{CMR2} the theory is compatible with cutting and gluing. In particular, the
composition in figure~\ref{gluing_moyal} yields the star product for the target Poisson structure.
\end{itemize}
Note that the associativity of Kontsevich's star product is now a consequence of the gluing formulae and of the associativity up to an $\Omega$-exact term for the $L_3$s. Having explicit formulae for some other cases, like the linear one, would be useful.

\begin{appendix}

\section{Computations of the mdQMEs of section \ref{mdQMEsection}} 
\label{appendA} 
 
\subsection{Computation for $L_3$}
\label{comp_L3}
We want to show the mdQME for $L_3$, i.e.
\begin{equation}
\label{mdQME_L3}
\nabla^{\partial L_3}_\textsf{G}\widehat{\psi}_{L_3}=\left(\dd+\I\hbar\triangle+\frac{\I}{\hbar}\Omega^{(3)}\right)\widehat{\psi}_{L_3}=0
\end{equation}
As it was already mentioned in section \ref{mdQMEsection}, \eqref{mdQME_L3} reduces to the equation $\Omega^{(3)}\widehat{\psi}_{L_3}=0$. Moreover, recall that $\Omega^{(3)}$ is given by (see the general construction of \cite{CMR2}) the sum of  
\begin{align}
\Omega^{(3)}_{\textnormal{pert}}&=-\frac{1}{2}\alpha^{ij}\left(\hbar^2\int_{\partial_1M}\frac{\delta}{\delta\mathbb{X}_i}\frac{\delta}{\delta \mathbb{X}_j}-\int_{\partial_2^{(1)}M\sqcup\partial_2^{(2)}M}\E_i\land\E_j\right)\\
\Omega^{(3)}_0&=\I\hbar\left(\int_{\partial_1M} \dd\mathbb{X}_i\frac{\delta}{\delta\mathbb{X}_i}+\int_{\partial_2^{(1)}M\sqcup\partial_2^{(2)}M}\dd\E_i\frac{\delta}{\delta\E_i}+\int_{\partial_1M}\dd x^{i}\frac{\delta}{\delta\mathbb{X}_i}+\int_{\partial_2M}\E_i \land\dd x^{i}\right),
\end{align}
since $\widehat{\boldsymbol{\eta}}=-\I\hbar\frac{\delta}{\delta \widehat{\textsf{X}}}$ is the conjugated momentum. Let us also look at some terms of $\Omega^{(3)}$ with a derivative by defining
\begin{align}
\Omega^{(3)}_{\mathbb{X},\textnormal{pert}}&:=-\frac{\hbar^2}{2}\alpha^{ij}\int_{\partial_1M}\frac{\delta}{\delta\mathbb{X}_i}\frac{\delta}{\delta\mathbb{X}_j},\\
\Omega_\mathbb{X}^{(3)}&:=\I\hbar\int_{\partial_1M}\dd\mathbb{X}_i\frac{\delta}{\delta\mathbb{X}_i},\\
\Omega_\mathbb{E}^{(3)}&:=\I\hbar\int_{\partial_2^{(1)}M\sqcup \partial_2^{(2)}M}\dd\E_i\frac{\delta}{\delta\mathbb{E}_i}
\end{align}

Applying $\widetilde{\Omega}=\Omega^{(3)}_{\textnormal{pert}}+\Omega^{(3)}_\mathbb{X}+\Omega^{(3)}_\E$ to the state $\widehat{\psi}_{L_3}$, we get
\[
\widetilde{\Omega}\widehat{\psi}_{L_3}=T_{L_3}\widetilde{\Omega}\ee^{\frac{\I}{\hbar}\mathcal{S}^{\textnormal{eff}}_{\partial M}}=T_{L_3}\left(\Omega^{(3)}_{\textnormal{pert}}+\Omega^{(3)}_\mathbb{X}+\Omega^{(3)}_\E\right)\ee^{\frac{\I}{\hbar}\mathcal{S}^{\textnormal{eff}}_{\partial M}}.
\]
Recall also that $T_{L_3}=1$. We want to compute each contribution of the different parts of $\widetilde\Omega$. Let us therefore first apply $\Omega^{(3)}_{\textnormal{pert}}$ and observe 

\begin{align}
\Omega^{(3)}_{\textnormal{pert}}\ee^{\frac{\I}{\hbar}\mathcal{S}^{\textnormal{eff}}_{\partial M}}&=-\frac{\hbar^2}{2}\alpha^{ij}\int_{\partial_1M}\frac{\delta}{\delta\mathbb{X}_i}\frac{\delta}{\delta\mathbb{X}_j}\ee^{\frac{\I}{\hbar}\mathcal{S}^{\textnormal{eff}}_{\partial M}}\\
&+\frac{1}{2}\alpha^{ij}\left(\int_{\partial_2^{(1)}M\sqcup\partial_2^{(2)}M}\E_i\land\E_j\right)\ee^{\frac{\I}{\hbar}\mathcal{S}^{\textnormal{eff}}_{\partial M}}\\
&=-\frac{\hbar^2}{2}\alpha^{ij}\left(\frac{\I}{\hbar}\right)\int_{\partial_1M} \frac{\delta}{\delta\mathbb{X}_i}\frac{\delta\mathcal{S}_{\partial M}^{\textnormal{eff}}}{\delta\mathbb{X}_j}\ee^{\frac{\I}{\hbar}\mathcal{S}^{\textnormal{eff}}_{\partial M}}\\
&+\frac{1}{2}\alpha^{ij}\left(\int_{\partial_2^{(1)}M\sqcup\partial_2^{(2)}M}\E_i\land\E_j\right)\ee^{\frac{\I}{\hbar}\mathcal{S}^{\textnormal{eff}}_{\partial M}}\\
&=-\frac{\hbar^2}{2}\alpha^{ij}\left(\frac{\I}{\hbar}\right)^2\int_{\partial_1M}\left(\frac{\delta^2\mathcal{S}_{\partial M}^{\textnormal{eff}}}{\delta\mathbb{X}_i\delta\mathbb{X}_j}\ee^{\frac{\I}{\hbar}\mathcal{S}^{\textnormal{eff}}_{\partial M}}+\frac{\delta\mathcal{S}_{\partial M}^{\textnormal{eff}}}{\delta\mathbb{X}_i}\frac{\delta\mathcal{S}_{\partial M}^{\textnormal{eff}}}{\delta\mathbb{X}_j}\ee^{\frac{\I}{\hbar}\mathcal{S}^{\textnormal{eff}}_{\partial M}}\right)\\
&+\frac{1}{2}\alpha^{ij}\left(\int_{\partial_2^{(1)}M\sqcup\partial_2^{(2)}M}\E_i\land\E_j\right)\ee^{\frac{\I}{\hbar}\mathcal{S}^{\textnormal{eff}}_{\partial M}}\\
\label{funcderiv}
&=-\frac{\hbar^2}{2}\alpha^{ij}\left(\frac{\I}{\hbar}\right)^2\int_{\partial_1M}\left(\frac{\delta^2\mathcal{S}_{\partial M}^{\textnormal{eff}}}{\delta\mathbb{X}_i\delta\mathbb{X}_j}+\frac{\delta\mathcal{S}_{\partial M}^{\textnormal{eff}}}{\delta\mathbb{X}_i}\frac{\delta\mathcal{S}_{\partial M}^{\textnormal{eff}}}{\delta\mathbb{X}_j}\right)\ee^{\frac{\I}{\hbar}\mathcal{S}^{\textnormal{eff}}_{\partial M}}\\
&+\frac{1}{2}\alpha^{ij}\left(\int_{\partial_2^{(1)}M\sqcup\partial_2^{(2)}M}\E_i\land\E_j\right)\ee^{\frac{\I}{\hbar}\mathcal{S}^{\textnormal{eff}}_{\partial M}}
\end{align}

Now we need to express the functional derivatives in (\ref{funcderiv}) in terms of the propagator and the fields. Each term is given by a sum and the only terms which contribute to the derivatives are 
\begin{align}
\frac{\delta^2\mathcal{S}_{\partial M}^{\textnormal{eff}}}{\delta\mathbb{X}_i\delta\mathbb{X}_j}&=\frac{\delta^2\mathcal{S}^{\textnormal{eff}}_{\partial^{(1)}M}}{\delta\mathbb{X}_i\delta\mathbb{X}_j}+\frac{\delta^2\mathcal{S}^{\textnormal{eff}}_{\partial^{(2)}M}}{\delta\mathbb{X}_i\delta\mathbb{X}_j}\\
\frac{\delta\mathcal{S}_{\partial M}^{\textnormal{eff}}}{\delta\mathbb{X}_i}&=\frac{\delta\mathcal{S}_{\partial^{(1)} M}^{\textnormal{eff}}}{\delta\mathbb{X}_i}+\frac{\delta\mathcal{S}_{\partial^{(2)} M}^{\textnormal{eff}}}{\delta\mathbb{X}_i}\\
\frac{\delta\mathcal{S}_{\partial M}^{\textnormal{eff}}}{\delta\mathbb{X}_j}&=\frac{\delta\mathcal{S}_{\partial^{(1)} M}^{\textnormal{eff}}}{\delta\mathbb{X}_j}+\frac{\delta\mathcal{S}_{\partial^{(2)} M}^{\textnormal{eff}}}{\delta\mathbb{X}_j},
\end{align}
since the other terms of the effective action do not depend on the $\mathbb{X}$-field. Now we get 
\begin{equation}
\frac{\delta^2\mathcal{S}^{\textnormal{eff}}_{\partial^{(k)}M}}{\delta\mathbb{X}_i\delta\mathbb{X}_j}=0,\hspace{0.3cm}
\frac{\delta\mathcal{S}_{\partial^{(k)} M}^{\textnormal{eff}}}{\delta\mathbb{X}_i}=\int_{\partial_1M\times\partial_{2}^{(k)}M}\zeta_{02}\land\pi_{2}^*\E_i,\hspace{0.3cm}
\frac{\delta\mathcal{S}_{\partial^{(k)} M}^{\textnormal{eff}}}{\delta\mathbb{X}_j}=\int_{\partial_1M\times\partial_{2}^{(k)}M}\zeta_{03}\land\pi_{2}^*\E_j
\end{equation}
and hence 
\begin{equation} 
\frac{\delta\mathcal{S}_{\partial M}^{\textnormal{eff}}}{\delta\mathbb{X}_i}\frac{\delta\mathcal{S}_{\partial M}^{\textnormal{eff}}}{\delta\mathbb{X}_j}=\frac{\delta\mathcal{S}_{\partial^{(1)} M}^{\textnormal{eff}}}{\delta\mathbb{X}_i}\frac{\delta\mathcal{S}_{\partial^{(1)} M}^{\textnormal{eff}}}{\delta\mathbb{X}_j}+\frac{\delta\mathcal{S}_{\partial^{(1)} M}^{\textnormal{eff}}}{\delta\mathbb{X}_i}\frac{\delta\mathcal{S}_{\partial^{(2)} M}^{\textnormal{eff}}}{\delta\mathbb{X}_j}+\frac{\delta\mathcal{S}_{\partial^{(2)} M}^{\textnormal{eff}}}{\delta\mathbb{X}_i}\frac{\delta\mathcal{S}_{\partial^{(1)} M}^{\textnormal{eff}}}{\delta\mathbb{X}_j}+\frac{\delta\mathcal{S}_{\partial^{(2)} M}^{\textnormal{eff}}}{\delta\mathbb{X}_i}\frac{\delta\mathcal{S}_{\partial^{(2)} M}^{\textnormal{eff}}}{\delta\mathbb{X}_j}.
\end{equation}

\begin{figure}[!ht]
\centering
\subfigure[For $\partial_2^{(1)}M$]{
\tikzset{
particle/.style={thick,draw=black},
particle2/.style={thick,draw=blue},
avector/.style={thick,draw=black, postaction={decorate},
    decoration={markings,mark=at position 1 with {\arrow[black]{triangle 45}}}},
gluon/.style={decorate, draw=black,
    decoration={coil,aspect=0}}
 }
\begin{tikzpicture}[x=0.04\textwidth, y=0.04\textwidth]
\draw[particle] (0,0)--(2,5) node[above]{};
\draw[particle2] (2,5)--(4,5) node[right]{};
\draw[particle2] (0,0)--(2,0) node[right]{};
\node[](u1) at (2,0){};
\node[](u2) at (4,0){};
\draw[particle2] (4,0)--(6,0) node[right]{};
\draw[particle] (6,0)--(4,5) node[right]{};
\node[](1) at (3,5){};
\draw[fill=black] (1) circle (0.05cm);
\node[](2) at (0.7,0){};
\node[](3) at (1.4,0){};
\node[](e1) at (1,-0.5){$\E^{(1)}$};
\node[](e2) at (5,-0.5){$\E^{(2)}$};
\node[](e3) at (3,5.5){$\mathbb{X}$};
\draw[avector] (3,5)--(0.7,0);
\draw[avector] (3,5)--(1.4,0);
\semiloop[particle]{u1}{u2}{0};
\end{tikzpicture}
}
\subfigure[For $\partial_2^{(2)}M$]{
\tikzset{
particle/.style={thick,draw=black},
particle2/.style={thick,draw=blue},
avector/.style={thick,draw=black, postaction={decorate},
    decoration={markings,mark=at position 1 with {\arrow[black]{triangle 45}}}},
gluon/.style={decorate, draw=black,
    decoration={coil,aspect=0}}
 }
\begin{tikzpicture}[x=0.04\textwidth, y=0.04\textwidth]
\draw[particle] (0,0)--(2,5) node[above]{};
\draw[particle2] (2,5)--(4,5) node[right]{};
\draw[particle2] (0,0)--(2,0) node[right]{};
\node[](u1) at (2,0){};
\node[](u2) at (4,0){};
\draw[particle2] (4,0)--(6,0) node[right]{};
\draw[particle] (6,0)--(4,5) node[right]{};
\node[](1) at (3,5){};
\draw[fill=black] (1) circle (0.05cm);
\node[](2) at (5.5,0){};
\node[](3) at (4.5,0){};
\node[](e1) at (1,-0.5){$\E^{(1)}$};
\node[](e2) at (5,-0.5){$\E^{(2)}$};
\node[](e3) at (3,5.5){$\mathbb{X}$};
\draw[avector] (3,5)--(5.5,0) node[right]{};
\draw[avector] (3,5)--(4.5,0) node[right]{};
\semiloop[particle]{u1}{u2}{0};
\end{tikzpicture}
}
\subfigure[For $\partial_{2}^{(1)}M\times\partial_2^{(2)}M$]{
\tikzset{
particle/.style={thick,draw=black},
particle2/.style={thick,draw=blue},
avector/.style={thick,draw=black, postaction={decorate},
    decoration={markings,mark=at position 1 with {\arrow[black]{triangle 45}}}},
gluon/.style={decorate, draw=black,
    decoration={coil,aspect=0}}
 }
\begin{tikzpicture}[x=0.04\textwidth, y=0.04\textwidth]
\draw[particle] (0,0)--(2,5) node[above]{};
\draw[particle2] (2,5)--(4,5) node[right]{};
\draw[particle2] (0,0)--(2,0) node[right]{};
\node[](u1) at (2,0){};
\node[](u2) at (4,0){};
\draw[particle2] (4,0)--(6,0) node[right]{};
\draw[particle] (6,0)--(4,5) node[right]{};
\node[](1) at (3,5){};
\draw[fill=black] (1) circle (0.05cm);
\node[](2) at (1,0){};
\node[](2) at (5,0){};
\node[](e1) at (1,-0.5){$\E^{(1)}$};
\node[](e2) at (5,-0.5){$\E^{(2)}$};
\node[](e3) at (3,5.5){$\mathbb{X}$};
\draw[avector] (3,5)--(5,0) node[right]{};
\draw[avector] (3,5)--(1,0) node[right]{};
\semiloop[particle]{u1}{u2}{0};
\end{tikzpicture}
}
\caption{The diagrams for the terms of $\Omega^{(3)}_{\textnormal{pert}}$.}
\label{op1}
\end{figure}
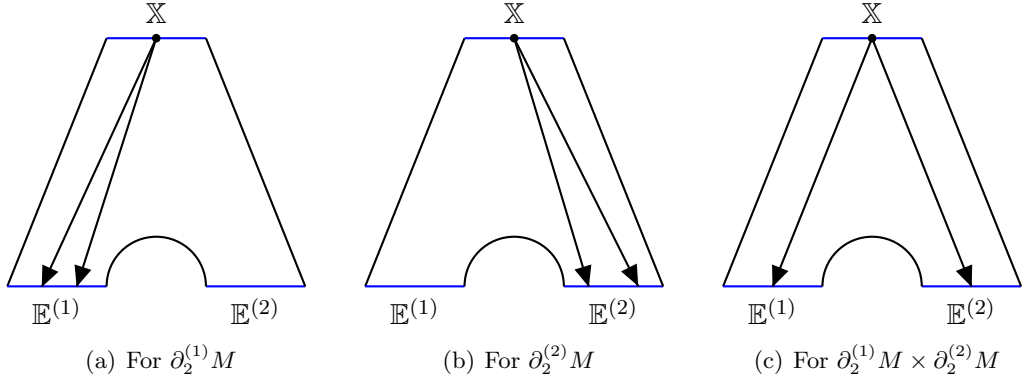

This shows that the application of $\Omega^{(3)}_{\textnormal{pert}}$ to the state $\widehat{\psi}_{L_3}$ gives

\begin{align}
\label{pert1}
\Omega^{(3)}_{\textnormal{pert}}\ee^{\frac{\I}{\hbar}\mathcal{S}^{\textnormal{eff}}_{\partial M}}
&=\frac{1}{2}\alpha^{ij}\left(\int_{\partial_1M\times C_2(\partial_2^{(1)}M)}\zeta_{02}\land\zeta_{03}\land\pi_{2,1}^*\E_i\land\pi_{2,2}^*\E_j\right)\ee^{\frac{\I}{\hbar}\mathcal{S}^{\textnormal{eff}}_{\partial M}}\\
\label{pert2}
&+\frac{1}{2}\alpha^{ij}\left(\int_{\partial_1M\times[\partial_2^{(1)}M\times\partial_2^{(2)}M]}\zeta_{02}\land\zeta_{03}\land\pi_{2,1}^*\E_i\land\pi_{2,2}^*\E_j\right)\ee^{\frac{\I}{\hbar}\mathcal{S}^{\textnormal{eff}}_{\partial M}}\\
\label{pert3}
&+\frac{1}{2}\alpha^{ij}\left(\int_{\partial_1M\times[\partial_2^{(2)}M\times\partial_2^{(1)}M]}\zeta_{02}\land\zeta_{03}\land\pi_{2,1}^*\E_i\land\pi_{2,2}^*\E_j\right)\ee^{\frac{\I}{\hbar}\mathcal{S}^{\textnormal{eff}}_{\partial M}}\\
\label{pert4}
&+\frac{1}{2}\alpha^{ij}\left(\int_{\partial_1M\times C_2(\partial_2^{(2)}M)}\zeta_{02}\land\zeta_{03}\land\pi_{2,1}^*\E_i\land\pi_{2,2}^*\E_j\right)\ee^{\frac{\I}{\hbar}\mathcal{S}^{\textnormal{eff}}_{\partial M}}\\
\label{pert5}
&+\frac{1}{2}\alpha^{ij}\left(\int_{\partial_2^{(1)}M\sqcup\partial_2^{(2)}M}\E_i\land\E_j\right)\ee^{\frac{\I}{\hbar}\mathcal{S}^{\textnormal{eff}}_{\partial M}}.
\end{align}
For these terms we get the diagrams as in figure \ref{op1}. Recall that $u_0\in\partial_1M,u_1\in M$ and $u_2,u_3,u_\nu\in\partial_2^{(k)}M$ for some $k\in\{1,2\}$. Now we want to compute the corresponding terms for $\Omega^{(3)}_\E$. We get
\begin{align}
\label{Eop1}
\Omega^{(3)}_\E\ee^{\frac{\I}{\hbar}\mathcal{S}^{\textnormal{eff}}_{\partial M}}
&=\I\hbar\left(\int_{\partial_2^{(1)}M\sqcup\partial_2^{(2)}M}\dd\E_i\frac{\delta\mathcal{S}^{\textnormal{eff}}_{\partial^{(1)}M}}{\delta\E_i}\right)\ee^{\frac{\I}{\hbar}\mathcal{S}^{\textnormal{eff}}_{\partial M}}\\
\label{Eop2}
&+\I\hbar\left(\int_{\partial_2^{(1)}M\sqcup\partial_2^{(2)}M}\dd\E_i\frac{\delta\mathcal{S}^{\textnormal{eff}}_{\partial^{(2)}M}}{\delta\E_i}\right)\ee^{\frac{\I}{\hbar}\mathcal{S}^{\textnormal{eff}}_{\partial M}}\\
\label{Eop3}
&+\I\hbar\left(\int_{\partial_2^{(1)}M\sqcup\partial_2^{(2)}M}\dd\E_i\frac{\delta\mathcal{S}^{\textnormal{pert,eff}}_{\partial_2^{(1)}M}}{\delta\E_i}\right)\ee^{\frac{\I}{\hbar}\mathcal{S}^{\textnormal{eff}}_{\partial M}}\\
\label{Eop4}
&+\I\hbar\left(\int_{\partial_2^{(1)}M\sqcup\partial_2^{(2)}M}\dd\E_i\frac{\delta\mathcal{S}^{\textnormal{pert,eff}}_{\partial_2^{(2)}M}}{\delta\E_i}\right)\ee^{\frac{\I}{\hbar}\mathcal{S}^{\textnormal{eff}}_{\partial M}}\\
\label{Eop5}
&+\I\hbar\left(\int_{\partial_2^{(1)}M\sqcup\partial_2^{(2)}M}\dd\E_i\frac{\delta\mathcal{S}^{\textnormal{pert,eff}}_{\partial_2^{(1)}M\sqcup\partial_2^{(2)}M}}{\delta\E_i}\right)\ee^{\frac{\I}{\hbar}\mathcal{S}^{\textnormal{eff}}_{\partial M}}
\end{align}

Let us now compute each term individually. We will start with (\ref{Eop1}) and observe
\begin{equation}
\I\hbar\left(\int_{\partial_2^{(1)}M\sqcup\partial_2^{(2)}M}\dd\E_i\frac{\delta\mathcal{S}^{\textnormal{eff}}_{\partial^{(1)}M}}{\delta\E_i}\right)\ee^{\frac{\I}{\hbar}\mathcal{S}^{\textnormal{eff}}_{\partial M}}
\label{Eop6} 
=\I\hbar\left(\frac{\I}{\hbar}\right)\left(-\int_{\partial_2^{(1)}M\times \partial_1M}\pi_1^*\dd\E_i\land\zeta_{0\nu}\land\pi_2^*\mathbb{X}_i\right)\ee^{\frac{\I}{\hbar}\mathcal{S}^{\textnormal{eff}}_{\partial M}}.
\end{equation}

For  (\ref{Eop2}) we get
\begin{equation}
\I\hbar\left(\int_{\partial_2^{(1)}M\sqcup\partial_2^{(2)}M}\dd\E_i\frac{\delta\mathcal{S}^{\textnormal{eff}}_{\partial^{(2)}M}}{\delta\E_i}\right)\ee^{\frac{\I}{\hbar}\mathcal{S}^{\textnormal{eff}}_{\partial M}}
\label{Eop7}
=\I\hbar\left(\frac{\I}{\hbar}\right)\left(-\int_{\partial_2^{(2)}M\times \partial_1M}\pi_1^*\dd\E_i\land\zeta_{0\nu}\land\pi_2^*\mathbb{X}_i\right)\ee^{\frac{\I}{\hbar}\mathcal{S}^{\textnormal{eff}}_{\partial M}}.
\end{equation}

For the term (\ref{Eop3}) we get
\begin{align}
&\I\hbar\left(\int_{\partial_2^{(1)}M\sqcup\partial_2^{(2)}M}\dd\E_i\frac{\delta\mathcal{S}^{\textnormal{pert,eff}}_{\partial_2^{(1)}M}}{\delta\E_i}\right)\ee^{\frac{\I}{\hbar}\mathcal{S}^{\textnormal{eff}}_{\partial M}}\\
&=\I\hbar\left(\frac{\I}{\hbar}\right)\frac{1}{2}\alpha^{ij}\left(\int_{C_2(\partial_2^{(1)}M)}\pi_1^*\dd\E_i\land\widehat{\zeta}^\partial_{23}\land\pi_2^*\E_j+\int_{C_2(\partial_2^{(1)}M)}\pi_1^*\E_i\land\widehat{\zeta}^\partial_{23}\land \pi_2^*\dd\E_j\right)\ee^{\frac{\I}{\hbar}\mathcal{S}^{\textnormal{eff}}_{\partial M}}.
\end{align}

Now using integration by parts we get 

\begin{align}
&\I\hbar\left(\frac{\I}{\hbar}\right)\frac{1}{2}\alpha^{ij}\left(\int_{C_2(\partial_2^{(1)}M)}\pi_1^*\dd\E_i\land\widehat{\zeta}^\partial_{23}\land\pi_2^*\E_j+\int_{C_2(\partial_2^{(1)}M)}\pi_1^*\E_i\land\widehat{\zeta}^\partial_{23}\land \pi_2^*\dd\E_j\right)\ee^{\frac{\I}{\hbar}\mathcal{S}^{\textnormal{eff}}_{\partial M}}\\
&=\I\hbar\left(\frac{\I}{\hbar}\right)\frac{1}{2}\alpha^{ij}\left(\int_{C_2(\partial_2^{(1)}M)}\dd(\pi_1^*\E_i\land\pi_2^*\E_j)\land\widehat{\zeta}^\partial_{23}\right)\ee^{\frac{\I}{\hbar}\mathcal{S}^{\textnormal{eff}}_{\partial M}}\\
\label{intparts1}
&=\I\hbar\left(\frac{\I}{\hbar}\right)\frac{1}{2}\alpha^{ij}\left(\int_{\partial_2^{(1)}M}\E_i\land\E_j\right)\ee^{\frac{\I}{\hbar}\mathcal{S}^{\textnormal{eff}}_{\partial M}}\\
\label{intparts2}
&+\I\hbar\left(\frac{\I}{\hbar}\right)\frac{1}{2}\alpha^{ij}\left(\int_{C_2(\partial_2^{(1)}M)}\pi_1^*\E_i\land\pi_2^*\E_j\land \dd\widehat{\zeta}^\partial_{23}\right)\ee^{\frac{\I}{\hbar}\mathcal{S}^{\textnormal{eff}}_{\partial M}}.
\end{align}

Because of the fact that we have vanishing cohomology, we can use Stokes' theorem to compute $\dd\widehat{\zeta}^\partial$. We get
\[
\dd\widehat{\zeta}^\partial_{23}:=\dd\widehat{\zeta}^\partial(u_2,u_3)=\int_{\partial_1M}\zeta_{02}\land\zeta_{03}.
\]
Hence (\ref{intparts2}) can be written as
\begin{equation}
\label{intparts5}
\I\hbar\left(\frac{\I}{\hbar}\right)\frac{1}{2}\alpha^{ij}\left(\int_{\partial_1M\times C_2(\partial_2^{(1)}M)}\zeta_{02}\land\zeta_{03}\land\pi_{2,1}^*\E_i\land\pi_{2,2}^*\E_j\right)\ee^{\frac{\I}{\hbar}\mathcal{S}^{\textnormal{eff}}_{\partial M}}.
\end{equation}

The same procedure holds for (\ref{Eop4}) and thus
\begin{align}
&\I\hbar\left(\int_{\partial_2^{(1)}M\sqcup\partial_2^{(2)}M}\dd\E_i\frac{\delta\mathcal{S}^{\textnormal{pert,eff}}_{\partial_2^{(2)}M}}{\delta\E_i}\right)\ee^{\frac{\I}{\hbar}\mathcal{S}^{\textnormal{eff}}_{\partial M}}\\
\label{intparts8}
&=\I\hbar\left(\frac{\I}{\hbar}\right)\frac{1}{2}\alpha^{ij}\left(\int_{\partial_2^{(2)}M}\E_i\land\E_j\right)\ee^{\frac{\I}{\hbar}\mathcal{S}^{\textnormal{eff}}_{\partial M}}\\
\label{intparts7}
&+\I\hbar\left(\frac{\I}{\hbar}\right)\frac{1}{2}\alpha^{ij}\left(\int_{\partial_1M\times C_2(\partial_2^{(2)}M)}\zeta_{02}\land\zeta_{03}\land\pi_{2,1}^*\E_i\land\pi_{2,2}^*\E_j\right)\ee^{\frac{\I}{\hbar}\mathcal{S}^{\textnormal{eff}}_{\partial M}}
\end{align}

The boundary propagator $\widehat{\zeta}^\partial$ is no longer in (\ref{intparts1}) and (\ref{intparts2}) since we have to integrate over the fiber of the configuration space, which implies that by the property of $\widehat{\zeta}^\partial$ its value is constant $1$ on the fiber. Moreover, since the diagonal is a copy of the manifold itself, we get that integration over $\partial C_2(\partial_2^{(k)}M)$ is then actually given by integration over $\partial_2^{(k)}M$ with the remaining form $\E_i\land\E_j$, i.e. evaluated at the same point for $k\in\{1,2\}$. For the term (\ref{Eop5}) we have the same principle and thus

\begin{align}
&\I\hbar\left(\int_{\partial_2^{(1)}M\sqcup\partial_2^{(2)}M}\dd\E_i\frac{\delta\mathcal{S}^{\textnormal{pert,eff}}_{\partial_2^{(1)}M\sqcup\partial_2^{(2)}M}}{\delta\E_i}\right)\ee^{\frac{\I}{\hbar}\mathcal{S}^{\textnormal{eff}}_{\partial M}}\\
\label{mixed}
&=\I\hbar\left(\frac{\I}{\hbar}\right)\frac{1}{2}\alpha^{ij}\left(\int_{\partial [\partial_2^{(1)}M\times\partial_2^{(2)}M]}\E_i\land\E_j\right)\ee^{\frac{\I}{\hbar}\mathcal{S}^{\textnormal{eff}}_{\partial M}}\\
\label{intparts6}
&+\I\hbar\left(\frac{\I}{\hbar}\right)\frac{1}{2}\alpha^{ij}\left(\int_{\partial_1M\times [\partial_2^{(1)}M\times\partial_2^{(2)}M]}\zeta_{02}\land\zeta_{03}\land\pi_{2,1}^*\E_i\land\pi_{2,2}^*\E_j\right)\ee^{\frac{\I}{\hbar}\mathcal{S}^{\textnormal{eff}}_{\partial M}}.
\end{align}

Moreover, we can observe that (\ref{mixed}) vanishes, since we get integration over the double boundary and the fields vanish on the endpoints of the boundary. Now we need to compute the terms for $\Omega^{(3)}_\mathbb{X}$. Therefore we get
\begin{align}
\label{Xop1}
\Omega_{\mathbb{X}}\ee^{\frac{\I}{\hbar}\mathcal{S}^{\textnormal{eff}}_{\partial M}}
&=\I\hbar\left(\frac{\I}{\hbar}\right)\left(\int_{\partial_1M}\dd\mathbb{X}_i\frac{\delta\mathcal{S}_{\partial^{(1)}M}^{\textnormal{eff}}}{\delta\mathbb{X}_i}\right)\ee^{\frac{\I}{\hbar}\mathcal{S}^{\textnormal{eff}}_{\partial M}}\\
\label{Xop2}
&+\I\hbar\left(\frac{\I}{\hbar}\right)\left(\int_{\partial_1M}\dd\mathbb{X}_i\frac{\delta\mathcal{S}_{\partial^{(2)}M}^{\textnormal{eff}}}{\delta\mathbb{X}_i}\right)\ee^{\frac{\I}{\hbar}\mathcal{S}^{\textnormal{eff}}_{\partial M}}.
\end{align}

The term in (\ref{Xop1}) is then 
\begin{equation} 
\label{Xop3}
\I\hbar\left(\frac{\I}{\hbar}\right)\left(-\int_{\partial_2^{(1)}M\times\partial_1M}\pi_1^*\E_i\land\zeta_{0\nu}\land \pi_2^*\dd\mathbb{X}_i\right)\ee^{\frac{\I}{\hbar}\mathcal{S}^{\textnormal{eff}}_{\partial M}}.
\end{equation}

The term in (\ref{Xop2}) is then 
\begin{equation}
\label{Xop4}
\I\hbar\left(\frac{\I}{\hbar}\right)\left(-\int_{\partial_2^{(2)}M\times\partial_1M}\pi_1^*\E_i\land\zeta_{0\nu}\land \pi_2^*\dd\mathbb{X}_i\right)\ee^{\frac{\I}{\hbar}\mathcal{S}^{\textnormal{eff}}_{\partial M}}.
\end{equation}

Now if we combine (\ref{Xop3}) with (\ref{Eop6}) and (\ref{Xop4}) with (\ref{Eop7}) and using again integration by parts we get 
\begin{align}
&\left(\int_{\partial_2^{(1)}M\times \partial_1M}\pi_1^*\dd\E_i\land\zeta_{0\nu}\land\pi_2^*\mathbb{X}_i+\int_{\partial_2^{(1)}M\times\partial_1M}\pi_1^*\E_i\land\zeta_{0\nu}\land \pi_2^*\dd\mathbb{X}_i\right)\ee^{\frac{\I}{\hbar}\mathcal{S}^{\textnormal{eff}}_{\partial M}}\\
&=\left(\int_{\partial_2^{(1)}M\times\partial_1M}\dd(\pi_1^*\E_i\land\pi_2^*\mathbb{X}_i)\land\zeta_{0\nu}\right)\ee^{\frac{\I}{\hbar}\mathcal{S}^{\textnormal{eff}}_{\partial M}}\\
\label{mixed2}
&=\left(\int_{\partial[\partial_2^{(1)}M\times\partial_1M]}\pi_1^*\E_i\land\pi_2^*\mathbb{X}_i\right)\ee^{\frac{\I}{\hbar}\mathcal{S}^{\textnormal{eff}}_{\partial M}}\\
\label{intparts3}
&+\left(\int_{\partial_2^{(1)}M\times\partial_1M}\pi_1^*\E_i\land \dd\zeta_{0\nu}\land\pi_2^*\mathbb{X}_i\right)\ee^{\frac{\I}{\hbar}\mathcal{S}^{\textnormal{eff}}_{\partial M}}
\end{align}

and

\begin{align}
&\left(\int_{\partial_2^{(2)}M\times \partial_1M}\pi_1^*\dd\E_i\land\zeta_{0\nu}\land\pi_2^*\mathbb{X}_i+\int_{\partial_2^{(2)}M\times\partial_1M}\pi_1^*\E_i\land\zeta_{0\nu}\land \pi_2^*\dd\mathbb{X}_i\right)\ee^{\frac{\I}{\hbar}\mathcal{S}^{\textnormal{eff}}_{\partial M}}\\
&=\left(\int_{\partial_2^{(2)}M\times\partial_1M}\dd(\pi_1^*\E_i\land\pi_2^*\mathbb{X}_i)\land\zeta_{0\nu}\right)\ee^{\frac{\I}{\hbar}\mathcal{S}^{\textnormal{eff}}_{\partial M}}\\
\label{mixed3}
&=\left(\int_{\partial[\partial_2^{(2)}M\times\partial_1M]}\pi_1^*\E_i\land\pi_2^*\mathbb{X}_i\right)\ee^{\frac{\I}{\hbar}\mathcal{S}^{\textnormal{eff}}_{\partial M}}\\
\label{intparts4}
&+\left(\int_{\partial_2^{(2)}M\times\partial_1M}\pi_1^*\E_i\land\dd\zeta_{0\nu}\land\pi_2^*\mathbb{X}_i\right)\ee^{\frac{\I}{\hbar}\mathcal{S}^{\textnormal{eff}}_{\partial M}}
\end{align}

respectively. Now again we can use that there is no cohomology and which implies that $\dd\zeta=0$ and thus the terms (\ref{intparts3}) and (\ref{intparts4}) vanish. Moreover, the terms (\ref{mixed2}) and (\ref{mixed3}) also vanish because of the principle we already had before. Now the term (\ref{pert1}) cancels with (\ref{intparts5}), the term (\ref{pert2}) cancels with (\ref{intparts6}) the term (\ref{pert4}) cancels with (\ref{intparts7}) and finally the term in (\ref{pert5}) cancels with the sum of the terms (\ref{intparts1}) and (\ref{intparts8}). Finally, for $\int_{\partial_1M}\dd x^{i}\frac{\delta}{\delta\mathbb{X}_i}$ we get a term $-\int_{\partial_2M\times M}\E_i\land\zeta_{0\nu}\land\dd x^{i}=-\int_{\partial_2M}\E_i\land \dd x^{i}$ which cancels the multiplicative term in $\Omega^{(3)}_0$. Now since $\dd\widehat{\psi}_{L_3}=0$, the mdQME holds for $\widehat{\psi}_{L_3}$, because $\triangle=0$ without cohomology.

\subsection{Computation for $L_1$}
\label{comp_L1}
Now we need to do the same computations for $M=L_1$ but with the difference that we have cohomology which means that $\triangle\not=0$. We need to show the mdQME for $L_1$, i.e.
\begin{equation}
\label{mdQME_L1}
\nabla_\textsf{G}^{\partial L_1}\widehat{\psi}_{L_1}^{\frac{\delta}{\delta\mathbb{X}}}=\left(\dd+\I\hbar\triangle+\frac{\I}{\hbar}\Omega^{(1)}_{\frac{\delta}{\delta\mathbb{X}}}+\right)\widehat{\psi}^{\frac{\delta}{\delta\mathbb{X}}}_{L_1}=0,
\end{equation}
where 
\begin{align}
\Omega^{(1)}_{\frac{\delta}{\delta\mathbb{X}}}&=\int_{\partial_2M}\left(\I\hbar \dd\E_i\frac{\delta}{\delta\E_i}-\hbar^2\E_i \land \dd x^{i}+\frac{1}{2}\alpha^{ij}\E_i\land\E_j\right),\\
\triangle&=\sum_{i=1}^n(-1)^{1+\deg z_i}\frac{\partial}{\partial z_i}\frac{\partial}{\partial z^{\dagger}_i}
\end{align}

We can use the formula $\deg z_i=1-\deg\chi_i$, where $\{[\chi_i]\}$ is a basis for the cohomology for some representatives $\chi_i$ (see \cite{CMR2}), and since we have the cohomology of the disk, we get that $\deg\chi_i=0$ and hence $\deg z_i=1$. Therefore we have an even exponent and only the coefficients $+1$. Now define $\Omega^{(1)}_\E:=\I\hbar\int_{\partial_2M}\dd\E_i\frac{\delta}{\delta\E_i}$. Then we get 

\begin{align}
\Omega^{(1)}_{\E}\widehat{\psi}^{\frac{\delta}{\delta\mathbb{X}}}_{L_1}&=\Omega^{(1)}_\E T_{L_1}\ee^{\frac{\I}{\hbar}\left(\mathcal{S}^{\textnormal{eff}}_{\partial M}\right)}=T_{L_1}\Omega^{(1)}_{\E}\ee^{\frac{\I}{\hbar}\left(\mathcal{S}^{\textnormal{pert,eff}}_{\partial_2 M}+\mathcal{S}^{\textnormal{eff}}_{\partial_2M}+\tilde{\mathcal{S}}^{\textnormal{pert,eff}}_{\partial_2M}\right)}\\
&=T_{L_1}\left(\I\hbar\int_{\partial_2M}\dd\E_i\frac{\delta}{\delta\E_i}\ee^{\frac{\I}{\hbar}\left(\mathcal{S}^{\textnormal{pert,eff}}_{\partial_2 M}+\mathcal{S}^{\textnormal{eff}}_{\partial_2M}+\tilde{\mathcal{S}}^{\textnormal{pert,eff}}_{\partial_2M}\right)}\right)\\
&=T_{L_1}\I\hbar\left(\frac{\I}{\hbar}\right)\int_{\partial_2M}\left( \dd\E_i\frac{\delta \mathcal{S}^{\textnormal{pert,eff}}_{\partial_2 M}}{\delta \E_i}+\dd\E_i\frac{\delta \mathcal{S}^{\textnormal{eff}}_{\partial_2M}}{\delta \E_i}+\dd\E_i\frac{\delta \mathcal{S}^{\textnormal{pert,eff}}_{\partial_2 M}}{\delta\E_i} \right)\ee^{\frac{\I}{\hbar}\mathcal{S}^{\textnormal{eff}}_{\partial M}}\\
\label{intparts_L1_1}
&=T_{L_1}\I\hbar\left(\frac{\I}{\hbar}\right)\frac{1}{2}\alpha^{ij}\left(\int_{C_2(\partial_2 M)} \pi_1^*\dd\E_i\land\widehat{\zeta}^\partial_{23}\land \pi_2^*\E_j\right)\ee^{\frac{\I}{\hbar}\mathcal{S}^{\textnormal{eff}}_{\partial M}}\\
\label{intparts_L1_2}
&+T_{L_1}\I\hbar\left(\frac{\I}{\hbar}\right)\frac{1}{2}\alpha^{ij}\left(\int_{C_2(\partial_2 M)} \pi_1^*\E_i\land\widehat{\zeta}^\partial_{23}\land \pi_2^*\dd\E_j\right)\ee^{\frac{\I}{\hbar}\mathcal{S}^{\textnormal{eff}}_{\partial M}}\\
\label{term_del_1}
&+T_{L_1}\I\hbar\left(\frac{\I}{\hbar}\right)\left(\int_{\partial_2M}z^{i}\dd\E_i\right)\ee^{\frac{\I}{\hbar}\mathcal{S}^{\textnormal{eff}}_{\partial M}}\\
\label{term_del_2}
&+T_{L_1}\I\hbar\left(\frac{\I}{\hbar}\right)\alpha^{ij}\left(\int_{\partial_2M}z_i^\dagger \dd\E_j\land\tau\right)\ee^{\frac{\I}{\hbar}\mathcal{S}^{\textnormal{eff}}_{\partial M}}
\end{align}

Again, with integration by parts, we get that (\ref{intparts_L1_1}) together with (\ref{intparts_L1_2}) gives
\begin{equation}
\label{intparts_L1_3}
T_{L_1}\I\hbar\left(\frac{\I}{\hbar}\right)\frac{1}{2}\alpha^{ij}\left(\int_{\partial_2M}\E_i\land\E_j\right)\ee^{\frac{\I}{\hbar}\mathcal{S}^{\textnormal{eff}}_{\partial M}}
+T_{L_1}\I\hbar\left(\frac{\I}{\hbar}\right)\frac{1}{2}\alpha^{ij}\left(\int_{C_2(\partial_2M)}\pi_1^*\E_i\land \dd\widehat{\zeta}^\partial_{23}\land \pi_2^*\E_j\right)\ee^{\frac{\I}{\hbar}\mathcal{S}^{\textnormal{eff}}_{\partial M}}.
\end{equation}
Now we can use (\ref{propagator1}) and we get that the second term of (\ref{intparts_L1_3}) is given by 
\begin{equation}
\label{intparts_L1_4}
T_{L_1}\I\hbar\left(\frac{\I}{\hbar}\right)\frac{1}{2}\alpha^{ij}\left(\int_{C_2(\partial_2M)}\pi_1^*\E_i\land\pi_1^*\tau\land\pi_2^*\E_j-\int_{C_2(\partial_2M)}\pi_1^*\E_i\land\pi_2^*\tau\land\pi_2^*\E_j\right)\ee^{\frac{\I}{\hbar}\mathcal{S}^{\textnormal{eff}}_{\partial M}}.
\end{equation}
Hence the term which arises from $\triangle$ cancels with (\ref{intparts_L1_4}). Moreover, we also get a term $\hbar^2\int_{\partial_2M}\E_i \land\dd x^{i}$, which cancels with the first multiplicative term of $\Omega^{(1)}_{\frac{\delta}{\delta\mathbb{X}}}$, and since (\ref{term_del_1}) and (\ref{term_del_2}) vanish, and clearly $\dd\widehat{\psi}^{\frac{\delta}{\delta\mathbb{X}}}_{L_1}=0$, we get that the \eqref{mdQME_L1} holds. Recall that the mdQME for $\widehat{\psi}_{L_1}^{\frac{\delta}{\delta\E}}$ is trivially satisfied, and $T_{L_1}=1$.

\section{Computations for the associativity of the gluing of section \ref{associativity}}
\label{comp_assoc}
We want to show \eqref{time_vary}. 
We claim that 
\begin{align}
\partial_t\widehat{\psi}_{\mathcal{M}^n}(t)&=\Omega^{(3)}\left(\widehat{\psi}_{\mathcal{M}^n}(t)\int_{\partial_2\mathcal{M}\times \partial_1\mathcal{M}}\pi_1^*\E_i\land\kappa^n\land\pi_2^*\mathbb{X}_i\right)\\
&+\Omega^{(3)}\left(\widehat{\psi}_{\mathcal{M}^n}(t)\frac{1}{2}\alpha^{ij}\int_{\mathcal{M}\times C_2(\partial_2\mathcal{M})}\zeta^{n,t}_{12}\land\kappa^{n}_{13}\land\pi_{2,1}^*\E_i\land\pi^*_{2,2}\E_j\right)\\
&+\Omega^{(3)}\left(\widehat{\psi}_{\mathcal{M}^n}(t)\frac{1}{2}\alpha^{ij}\int_{\mathcal{M}\times C_2(\partial_2\mathcal{M})}\kappa_{12}^n\land\zeta^{n,t}_{13}\land\pi_{2,1}^*\E_i\land\pi_{2,2}^*\E_j\right)\\
\label{claim}
&=\Omega^{(3)}\left(\widehat{\psi}_{\mathcal{M}^n}(t)\A\right)
\end{align}
with 
\begin{multline*}
\A=\int_{\partial_2\mathcal{M}\times \partial_1\mathcal{M}}\pi^*_1\E_i\land \kappa^n\land \pi_2^*\mathbb{X}_i+\frac{1}{2}\alpha^{ij}\int_{\mathcal{M}\times C_2(\partial_2\mathcal{M})}\zeta_{12}^{n,t}\land\kappa^n_{13}\land\pi_{2,1}^*\E_i\land\pi_{2,2}^*\E_j\\+\frac{1}{2}\alpha^{ij}\int_{\mathcal{M}\times C_2(\partial_2\mathcal{M})}\kappa_{12}^n\land\zeta_{13}^{n,t}\land\pi_{2,1}^*\E_i\land\pi_{2,2}^*\E_j,
\end{multline*}
Indeed, we can first observe that 
\begin{equation}
\partial_t\widehat{\psi}_{\mathcal{M}^n}(t)=\left(\frac{\I}{\hbar}\right)\widehat{\psi}_{\mathcal{M}^n}(t)(\partial_t\mathcal{S}_{\mathcal{M}^n}^{\textnormal{eff}}(t)),
\end{equation}
This shows that we only have to compute $\partial_t\mathcal{S}_{\mathcal{M}^n}^{\textnormal{eff}}(t)$. Let us first look at the free term $\mathcal{S}_{\mathcal{M}^n}^{\textnormal{free,eff}}(t)$ of the action. We get that its derivative is given by
\begin{equation}
\label{free1}
\partial_t\mathcal{S}_{\mathcal{M}^n}^{\textnormal{free,eff}}(t)=\partial_t\left(-\int_{\partial_2\mathcal{M}\times\partial_1\mathcal{M}}\pi_1^*\E_i\land(\zeta+t\dd\kappa^n)\land\pi_2^*\mathbb{X}_i\right)=-\int_{\partial_2\mathcal{M}\times\partial_1\mathcal{M}}\pi_1^*\E_i\land \dd\kappa^n\land\pi_2^*\mathbb{X}_i.
\end{equation}
The derivative corresponding to the perturabation term $\mathcal{S}_{\mathcal{M}^n}^{\textnormal{pert,eff}}(t)$ is given by 
\begin{equation}
\label{pert111}
\partial_t\mathcal{S}_{\mathcal{M}^n}^{\textnormal{pert,eff}}(t)=\alpha^{ij}\int_{\mathcal{M}\times C_2(\partial_2\mathcal{M})}\zeta^{n,t}_{12}\land\partial_t\zeta^{n,t}_{13}\land\pi_{2,1}^*\E_i\land\pi_{2,2}^*\E_j=\alpha^{ij}\int_{\mathcal{M}\times C_2(\partial_2\mathcal{M})}\zeta^{n,t}_{12}\land \dd\kappa^n_{13}\land\pi_{2,1}^*\E_i\land\pi_{2,2}^*\E_j.
\end{equation}

Hence we get that 
\begin{equation}
\label{state_der}
\partial_t\widehat{\psi}_{\mathcal{M}^n}(t)=\left(\frac{\I}{\hbar}\right)\widehat{\psi}_{\mathcal{M}^n}(t)\left(-\int_{\partial_2\mathcal{M}\times\partial_1\mathcal{M}}\pi_1^*\E_i\land \dd\kappa^n\land\pi_2^*\mathbb{X}_i+\alpha^{ij}\int_{\mathcal{M}\times C_2(\partial_2\mathcal{M})}\zeta^{n,t}_{12}\land \dd\kappa^n_{13}\land\pi_{2,1}^*\E_i\land\pi_{2,2}^*\E_j\right).
\end{equation}

Now we want to compute $\Omega^{(3)}\left(\widehat{\psi}_{\mathcal{M}^n}(t)\A\right)$. We get 
\begin{align}
\Omega^{(3)}(\widehat{\psi}_{\mathcal{M}^n}(t)\A)&=\left(\Omega^{(3)}_{0}+\Omega^{(3)}_{\textnormal{pert}}\right)\left(\widehat{\psi}_{\mathcal{M}^n}(t)\A\right)\\
&=\Omega^{(3)}_0\left(\widehat{\psi}_{\mathcal{M}^n}(t)\A\right)+\Omega^{(3)}_{\textnormal{pert}}\left(\widehat{\psi}_{\mathcal{M}^n}(t)\A\right)\\
&=\underbrace{\left(\Omega^{(3)}_0\widehat{\psi}_{\mathcal{M}^n}(t)\right)\A}_{(\star)}+\underbrace{\left(\Omega^{(3)}_0\A\right)\widehat{\psi}_{\mathcal{M}^n}(t)}_{(\star\star)}+\underbrace{\Omega^{(3)}_{\textnormal{pert}}\left(\widehat{\psi}_{\mathcal{M}^n}(t)\A\right)}_{(\star\star\star)}.
\end{align}

Let us first compute term $(\star\star\star)$. We get 
\begin{equation}
\Omega^{(3)}_{\textnormal{pert}}\left(\widehat{\psi}_{\mathcal{M}^n}(t)\cdot\A\right)=-\frac{\hbar^2}{2}\alpha^{ij}\int_{\partial_1M}\frac{\delta}{\delta\mathbb{X}_i}\frac{\delta}{\delta\mathbb{X}_j}\left(\widehat{\psi}_{\mathcal{M}^n}(t)\A\right)+\frac{1}{2}\alpha^{ij}\int_{\partial_2M}\E_i\land\E_j\left(\widehat{\psi}_{\mathcal{M}^n}(t)\A\right)
\end{equation}
\begin{equation}
=-\frac{\hbar^2}{2}\alpha^{ij}\int_{\partial_1M}\frac{\delta}{\delta\mathbb{X}_i}\left[\left(\frac{\I}{\hbar}\right)\frac{\delta \mathcal{S}^{\textnormal{eff}}_{\mathcal{M}^n}}{\delta\mathbb{X}_j}\widehat{\psi}_{\mathcal{M}^n}(t)\A+\frac{\delta\A}{\delta\mathbb{X}_j}\widehat{\psi}_{\mathcal{M}^n}(t)\right]
+\frac{1}{2}\alpha^{ij}\int_{\partial_2M}\E_i\land\E_j\left(\widehat{\psi}_{\mathcal{M}^n}(t)\A\right)
\end{equation}
\begin{align}
=-\frac{\hbar^2}{2}\alpha^{ij}\left(\frac{\I}{\hbar}\right)^2\int_{\partial_1M}\frac{\delta\mathcal{S}^{\textnormal{eff}}_{\mathcal{M}^n}}{\delta\mathbb{X}_i}\frac{\delta\mathcal{S}^{\textnormal{eff}}_{\mathcal{M}^n}}{\delta\mathbb{X}_j}\widehat{\psi}_{\mathcal{M}^n}(t)\A\\
\label{term1}
&+\frac{\hbar^2}{2}\alpha^{ij}\left(\frac{\I}{\hbar}\right)\int_{\partial_1M}\frac{\delta\A}{\delta\mathbb{X}_i}\frac{\delta\mathcal{S}^{\textnormal{eff}}_{\mathcal{M}^n}}{\delta\mathbb{X}_j}\widehat{\psi}_{\mathcal{M}^n}(t)\\
\label{term2}
&+\frac{\hbar^2}{2}\alpha^{ij}\left(\frac{\I}{\hbar}\right)\int_{\partial_1M}\frac{\delta\A}{\delta\mathbb{X}_j}\frac{\delta\mathcal{S}^{\textnormal{eff}}_{\mathcal{M}^n}}{\delta\mathbb{X}_i}\widehat{\psi}_{\mathcal{M}^n}(t)\\
\label{term3}
&+\frac{1}{2}\alpha^{ij}\int_{\partial_2M}\E_i\land\E_j\left(\widehat{\psi}_{\mathcal{M}^n}(t)\A\right)
\end{align}

Analyzing the terms, we get that term (\ref{term1}) is given by 
\begin{equation}
-\frac{\hbar^2}{2}\alpha^{ij}\left(\int_{\partial_1\mathcal{M}\times C_2(\partial_2\mathcal{M})}\zeta_{12}^{n,t}\land\zeta_{13}^{n,t}\land\pi_{2,1}^*\E_i\land\pi_{2,2}^*\E_j\right)\A\widehat{\psi}_{\mathcal{M}^n}(t) ,
\end{equation}

term (\ref{term2}) by 
\begin{equation}
\label{term22}
\frac{\hbar^2}{2}\alpha^{ij}\left(\int_{\partial_1\mathcal{M}\times C_2(\partial_2\mathcal{M})}\zeta_{12}^{n,t}\land\kappa_{13}^n\land\pi_{2,1}^*\E_i\land\pi_{2,2}^*\E_j\right)\widehat{\psi}_{\mathcal{M}^n}(t),
\end{equation}

and term (\ref{term3}) by 

\begin{equation}
\label{term33}
\frac{\hbar^2}{2}\alpha^{ij}\left(\int_{\partial_1\mathcal{M}\times C_2(\partial_2\mathcal{M})}\zeta_{13}^{n,t}\land\kappa_{12}^n\land\pi_{2,1}^*\E_i\land\pi_{2,2}^*\E_j\right)\widehat{\psi}_{\mathcal{M}^n}(t),
\end{equation}

We can take the sum of (\ref{term22}) and (\ref{term33}) to obtain 
\begin{equation}
{\hbar^2}\alpha^{ij}\left(\int_{\partial_1\mathcal{M}\times C_2(\partial_2\mathcal{M})}\zeta_{12}^{n,t}\land\kappa_{13}^n\land\pi_{2,1}^*\E_i\land\pi_{2,2}^*\E_j\right)\widehat{\psi}_{\mathcal{M}^n}(t),
\end{equation}

Now we want to compute $(\star)$. We get 
\begin{align}
\left(\Omega^{(3)}_0\widehat{\psi}_{\mathcal{M}^n}(t)\right)\A&=\left(\frac{\I}{\hbar}\right)\left(\Omega^{(3)}_0\mathcal{S}^{\textnormal{eff}}_{\mathcal{M}^n}\right)\widehat{\psi}_{\mathcal{M}^n}(t)\A\\
&=-\I\hbar\left(\frac{\I}{\hbar}\right)\left(\overbrace{\int_{\partial_2\mathcal{M}\times\partial_1\mathcal{M}}\pi_{1}^*\dd\E_i\land\zeta_{12}^{n,t}\land\pi_{2}^*\mathbb{X}_i+\int_{\partial_2\mathcal{M}\times\partial_1\mathcal{M}}\pi_{1}^*\E_i\land\zeta_{12}^{n,t}\land\pi_{2}^*\dd\mathbb{X}_i}^{\text{integration by parts}=\int_{\partial_2\mathcal{M}\times\partial_1\mathcal{M}}\pi_1^*\E_i\land\zeta^{n,t}_{02}\land\pi_2^*\mathbb{X}_i=0}\right)\widehat{\psi}_{\mathcal{M}^n}(t)\A\\
&+\I\hbar\left(\frac{\I}{\hbar}\right)\left(\frac{1}{2}\alpha^{ij}\int_{\mathcal{M}\times C_2(\partial_2\mathcal{M})}\zeta_{12}^{n,t}\land\zeta_{13}^{n,t}\land\pi_1^*\dd\E_i\land\pi^*\E_j\right)\widehat{\psi}_{\mathcal{M}^n}(t)\A\\
&+\I\hbar\left(\frac{\I}{\hbar}\right)\left(\frac{1}{2}\alpha^{ij}\int_{\mathcal{M}\times C_2(\partial_2\mathcal{M})}\zeta_{12}^{n,t}\land\zeta_{13}^{n,t}\land\pi_1^*\E_i\land\pi^*\dd\E_j\right)\widehat{\psi}_{\mathcal{M}^n}(t)\A
\end{align}
\begin{equation}
=\I\hbar\left(\frac{\I}{\hbar}\right)\Bigg(\underbrace{\int_{\partial_1\mathcal{M}\times C_2(\partial_2\mathcal{M})}\dd(\zeta^{n,t}_{02}\land\zeta^{n,t}_{03})\land \pi_{2,1}^*\E_i\land\pi_{2,2}^*\E_j}_{=0}
+\left(\frac{1}{2}\alpha^{ij}\int_{\partial_2\mathcal{M}}\E_i\land\E_j\right)\widehat{\psi}_{\mathcal{M}^n}(t)\Bigg)\A.
\end{equation}

The term $(\star\star)$ gives us
\begin{align}
\left(\Omega^{(3)}_0\A\right)\widehat{\psi}_{\mathcal{M}^n}(t)&=\I\hbar\left(\int_{\partial_2\mathcal{M}\times \partial_1\mathcal{M}}\pi_1^*\dd\E_i\land\kappa^n_{01}\land\pi_2^*\mathbb{X}_i+\int_{\partial_2\mathcal{M}\times \partial_1\mathcal{M}}\pi_1^*\E_i\land\kappa^n_{01}\land\pi_2^*\dd\mathbb{X}_i\right)\widehat{\psi}_{\mathcal{M}^n}(t)\\
&+\I\hbar\left(\alpha^{ij}\int_{\mathcal{M}\times C_2(\partial_2\mathcal{M})}\zeta_{12}^{n,t}\land\kappa^n_{13}\land\pi_{2,1}^*\dd\E_i\land\pi_{2,2}^*\E_j\right)\widehat{\psi}_{\mathcal{M}^n}(t)\\
&+\I\hbar\left(\alpha^{ij}\int_{\mathcal{M}\times C_2(\partial_2\mathcal{M})}\zeta_{12}^{n,t}\land\kappa^n_{13}\land\pi_{2,1}^*\E_i\land\pi_{2,2}^*\dd\E_j\right)\widehat{\psi}_{\mathcal{M}^n}(t)\\
&=\I\hbar\left(\int_{\partial_2\mathcal{M}\times \partial_1\mathcal{M}}\pi_1^*\E_i\land \dd\kappa^n_{12}\land\pi_2^*\mathbb{X}_i\right)\widehat{\psi}_{\mathcal{M}^n}(t)\\
&+\I\hbar\left(\alpha^{ij}\int_{\partial_1\mathcal{M}\times C_2(\partial_2\mathcal{M})}\zeta_{02}^{n,t}\land\kappa^{n}_{03}\land\pi_{2,1}^*\E_i\land\pi_{2,2}^*\E_j\right)\widehat{\psi}_{\mathcal{M}^n}(t)\\
&+\I\hbar\left(\alpha^{ij}\int_{\mathcal{M}\times C_2(\partial_2\mathcal{M})}\zeta_{12}^{n,t}\land \dd\kappa^{n}_{13}\land\pi_{2,1}^*\E_i\land\pi_{2,2}^*\E_j\right)\widehat{\psi}_{\mathcal{M}^n}(t)
\end{align}

Rearranging the terms, and by the fact that $\widehat{\psi}_{\mathcal{M}^n}(t)$ satisfies the mdQME, we see that \eqref{time_vary} holds.
\end{appendix}

\end{document}